\documentclass{juliacon}
\usepackage{amsmath}
\usepackage{xspace}
\usepackage{booktabs}
\usepackage{upgreek}
\usepackage[most]{tcolorbox}
\tcbset{on line,boxsep=1pt,left=0pt,right=0pt,top=0pt,bottom=0pt,
        colframe=white,colback=gray!10,highlight math style={enhanced}}
\usepackage[labelformat=simple]{subfig}

\usepackage{tikz}

\newcommand{\code}[1]{\tcbox{\texttt{#1}}\xspace}
\newcommand{\binary}[1]{#1\ensuremath{^b}\xspace}
\newcommand{\R}{\ensuremath{\mathbb{R}}\xspace}
\newcommand{\X}{\ensuremath{X}\xspace}
\newcommand{\Y}{\ensuremath{Y}\xspace}
\newcommand{\Z}{\ensuremath{Z}\xspace}
\newcommand{\U}{\ensuremath{U}\xspace}
\newcommand{\CH}{\ensuremath{\mathit{CH}}\xspace}

\begin{document}

\title{LazySets.jl: Scalable Symbolic-Numeric Set Computations}

\author[1]{Marcelo Forets$^{*,}$}
\author[2]{Christian Schilling$^{*,}$}
\affil[1]{Universidad de la Rep\'{u}blica, Uruguay}
\affil[2]{Aalborg University, Denmark}


\hypersetup{
pdftitle = {LazySets.jl: Scalable Symbolic-Numeric Set Computations},
pdfsubject = {JuliaCon 2019 Proceedings},
pdfauthor = {Marcelo Forets, Christian Schilling},
pdfkeywords = {Set propagation, Geometry, Reachability analysis, Hybrid system, Support function, Formal verification},
}

\maketitle

\begin{abstract}
LazySets.jl is a Julia library that provides ways to symbolically represent sets of points as geometric shapes, with a special focus on convex sets and polyhedral approximations. LazySets provides methods to apply common set operations, convert between different set representations, and efficiently compute with sets in high dimensions using specialized algorithms based on the set types.
LazySets is the core library of JuliaReach, a cutting-edge software addressing the fundamental problem of reachability analysis: computing the set of states that are reachable by a dynamical system from all initial states and for all admissible inputs and parameters.
While the library was originally designed for reachability and formal verification, its scope goes beyond such topics.
LazySets is an easy-to-use, general-purpose and scalable library for computations that mix symbolics and numerics.
In this article we showcase the basic functionality, highlighting some of the key design choices.

\end{abstract}

\insert\footins{\noindent\footnotesize The published version of this paper can be found at \url{https://doi.org/10.21105/jcon.00097}.}

\renewcommand{\thefootnote}{\fnsymbol{footnote}}
\footnotetext[1]{Both authors contributed equally.}
\renewcommand{\thefootnote}{\arabic{footnote}}

\section{Introduction}

\begin{figure}[t]
	\centering
	\includegraphics[width=0.7\linewidth, keepaspectratio]{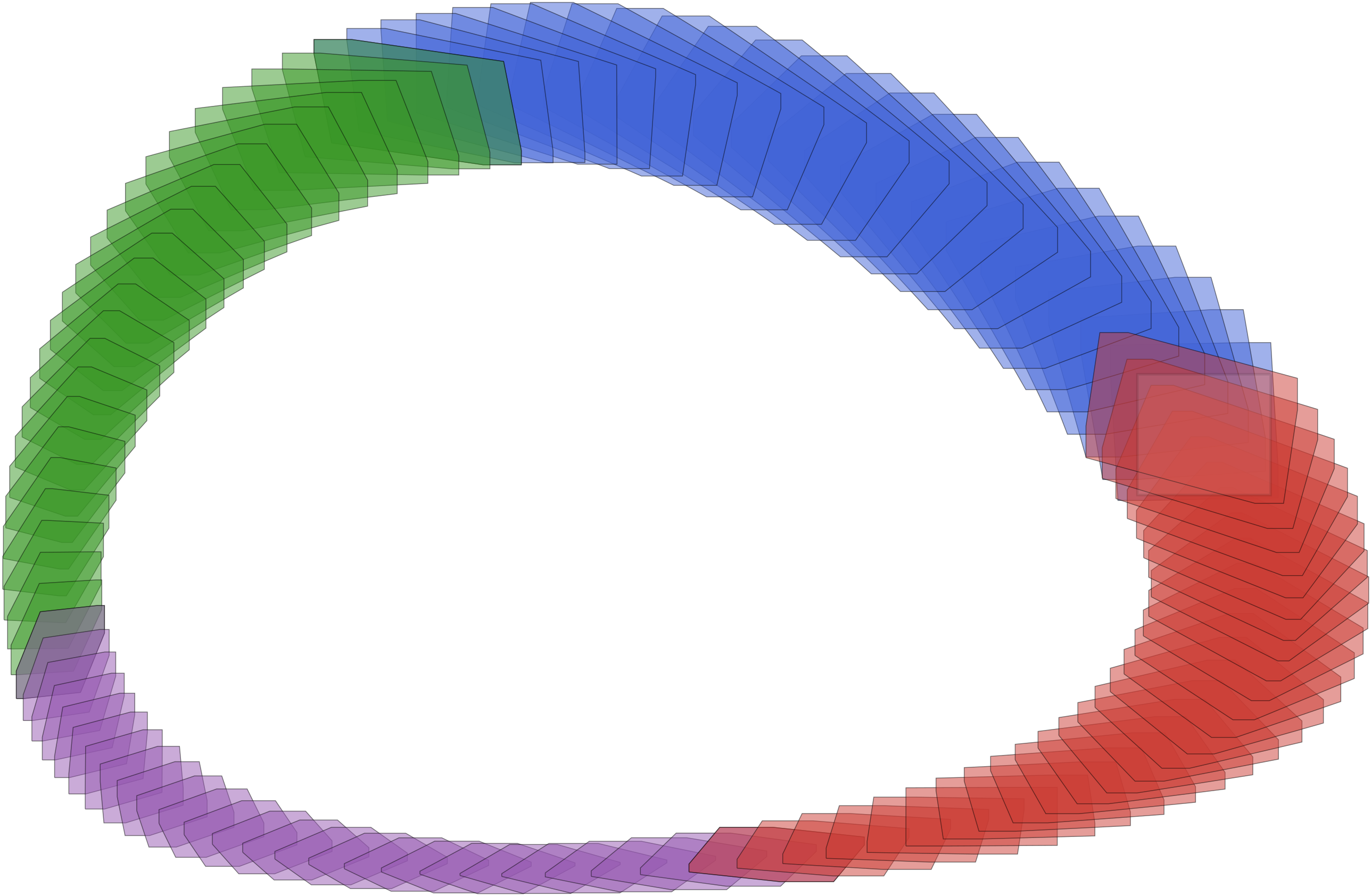}
	\vspace*{1mm}
	\caption{Example of set propagation using LazySets.
	Starting from a set of initial conditions represented as a hyperrectangle (on the right) and a given dynamical system (in this case, the Lotka-Volterra equations), a reachability algorithm computes a sequence of Taylor models. For plotting, the Taylor models are overapproximated using zonotopes.
	It is commonly required to use different set representations in reachability algorithms.}
	\label{fig:lotka_volterra}
\end{figure}

\href{https://github.com/JuliaReach/LazySets.jl}{LazySets.jl} is an open-source Julia package for calculus with geometric sets of points in Euclidean space. For a visual example applying LazySets, see Fig.~\ref{fig:lotka_volterra}.
(The set representations used there will be introduced later.)
The package provides solutions to represent sets, perform calculations on them, and combine them via set operations to form new sets.
A key aspect of LazySets is that set operations can be applied concretely, meaning that a computation is invoked, or lazily, meaning that the computation is delayed.
Based on geometric concepts, LazySets can evaluate queries on the lazy set representation, which enables efficient operation in very high dimensions that is not possible when applying the operations concretely.

\smallskip

LazySets aims to be a flexible and scalable library.
It provides specialized representations for various common classes of sets and ways for interacting with these sets, and is able to work with complex set constructs by use of the support-function calculus.\footnote{Complementary background is included in the Appendix.}
Flexibility is achieved by implementing generic algorithms that apply to multiple types of sets, and interoperability is achieved by connecting all set types through common interfaces.
Efficiency is achieved by adding special-case implementations where applicable; set operations are often binary functions and Julia's multiple dispatch greatly simplifies the choice of the most efficient implementation for a given combination of sets.
LazySets is designed to work with very high-dimensional sets but also provides specialized methods for one- and two-dimensional sets.
Finally, LazySets is well integrated with the Julia ecosystem for scientific computing~\cite{bezanson2017julia}, and additional functionality is available upon loading optional packages.
The target audience of LazySets are researchers who use symbolic and numeric set computations.

\smallskip

In this article we present the basic functionality of LazySets, starting with common sets and operations in Section~\ref{sec:basic}.
More advanced topics on type composition and conversion are introduced in sections \ref{sec:lazy} and \ref{sec:approx}.
Several applications are included in Section~\ref{sec:applications}.
We comment on related libraries in Section~\ref{sec:conclusion}.
Installation instructions can be found in Appendix~\ref{sec:installation}.
Background mathematical definitions are included in Appendix~\ref{sec:mathdef}.
The code to reproduce the figures can be found in Appendix~\ref{sec:code_examples}.

\section{Basic sets and operations} \label{sec:basic}

\begin{table*}[tb]
	\tbl{Set operations available in LazySets. The distinction between lazy and concrete functions is explained in Section~\ref{sec:lazy}. The result of the last three operations is generally not convex even if used with convex operands. For binary operations (marked with \binary{$\cdot$}) there is also an $n$-ary lazy version with the suffix \code{Array}, e.g., \code{MinkowskiSumArray}. Unicode symbols (as mentioned in the column ``Short form'') are entered in the Julia REPL by typing the \LaTeX\ command (e.g.: \code{\textbackslash{}oplus} for $\oplus$) followed by pressing the ``Tab'' key. See Appendix~\ref{sec:setops} for some central definitions.}{
	\begin{tabular}{l l l l l}
		\toprule
		Operation name & Math form & Lazy function (constructor) & Short form & Concrete function \\
		\midrule
		\binary{Minkowski sum} & $\X \oplus \Y$ & \code{MinkowskiSum} & \code{+},\code{\textbackslash{}oplus} & \code{minkowski$\_$sum} \\
		\binary{Intersection} & $\X \cap \Y$ & \code{Intersection} & \code{\textbackslash{}cap} & \code{intersection} \\
		\binary{Cartesian product} & $\X \times \Y$ & \code{CartesianProduct} & \code{*},\code{\textbackslash{}times} & \code{cartesian$\_$product} \\
		\binary{Convex hull} & $\CH(\X \cup \Y)$ & \code{ConvexHull} & \code{CH} & \code{convex$\_$hull} \\
		Symmetric interval hull & $\boxdot(\X)$ & \code{SymmetricIntervalHull} & \code{\textbackslash{}boxdot} & \code{symmetric$\_$interval$\_$hull} \\
		Linear map & $A \X$ & \code{LinearMap} & \code{*} & \code{linear$\_$map} \\
		Exponential map & $e^A \X$ & \code{ExponentialMap} && \code{exponential$\_$map} \\
		Translation & $\X + b$ & \code{Translation} & \code{+} & \code{translate} \\
		Affine map & $A \X + b$ & \code{AffineMap} & \code{*} and \code{+} & \code{affine$\_$map} \\
		Reset map & $x_i \mapsto c$ & \code{ResetMap} && - \\
		Inverse linear map & $A^{-1} \X$ & \code{InverseLinearMap} && - \\
		Bloating & $\X \oplus \{x : \Vert x \Vert \leq \varepsilon\}$ & \code{Bloating} && - \\
		\midrule
		\binary{Union} & $\X \cup \Y$ & \code{UnionSet} & \code{\textbackslash{}cup} & - \\
		Complement & $\X^C$ & \code{Complement} && \code{complement} \\
		Rectified linear unit & $x_i \mapsto \max(x_i, 0)$ & \code{Rectification} && \code{rectify} \\
		\bottomrule
	\end{tabular}}
	\label{tab:operations}
\end{table*}

LazySets offers support for \emph{convex} and \emph{non-convex} sets.
Intuitively, a set $\X$ is convex if one can draw a straight line segment between any two points in $\X$ without leaving $\X$ (see Appendix~\ref{sec:convexdef} for a formal definition).
This explains why optimization over a convex set is efficient.
Convex sets enjoy several other attractive properties, and many important geometric shapes are convex.

\subsection{Constructing sets}

Two basic sets are the \emph{hyperplane}
\[
	\{x \in \R^n \mid a^T x = b\},
\]
which is parametric in a vector $a \in \R^n$ and a scalar $b \in \R$, and the \emph{half-space} (or \emph{linear constraint})
\[
	\{x \in \R^n \mid a^T x \leq b\},
\]
which consists of all points on one side of the corresponding hyperplane.
In LazySets these sets are constructed from $a$ and $b$. For example, the two-dimensional hyperplane $x = 1$ (resp. the half-space $x \leq 1$) are:

\begin{minipage}{\linewidth}
\vspace{-\abovedisplayskip}
\begin{lstlisting}
julia> a = [1.0, 0.0]; b = 1.0

julia> Hyperplane(a, b)
Hyperplane{Float64,Vector{Float64}}([1.0, 0.0], 1.0)

julia> HalfSpace(a, b)
HalfSpace{Float64,Vector{Float64}}([1.0, 0.0], 1.0)
\end{lstlisting}
\end{minipage}
Higher-dimensional sets are defined in a similar fashion; for instance, the 100-dimensional half-space $x_1 + \ldots + x_{100} \leq 10$ is:\footnote{Sometimes we abbreviate the output to improve readability.}
\begin{minipage}{\linewidth}
\vspace{-\abovedisplayskip}
	\begin{lstlisting}
julia> a = fill(1.0, 100); b = 10.0

julia> HalfSpace(a, b)
HalfSpace([1.0, ..., 1.0], 10.0)
	\end{lstlisting}
\end{minipage}

The most widely used convex sets in various disciplines are (convex) \emph{polyhedra}, which are characterized as the finite intersection of half-spaces.
This is also known as the \emph{H-representation}.
For bounded polyhedra, which are called \emph{polytope}, LazySets also supports a dual way to represent such sets in \emph{V-representation} as the convex hull of points.
The functions \code{tohrep(X)} and \code{tovrep(X)} can be used to convert between these representations.

Optimization over a polyhedron with linear objective corresponds to solving a linear program.
Linear programs can model a wide variety of real-life optimization problems and can be solved efficiently \cite{dantzig1998linear,kochenderfer2019algorithms}.
On the left of Fig.~\ref{fig:supfunc} we show an example of a polytope in orange, with seven (linear) constraints, and a half-space in blue.

\smallskip

LazySets contains many (currently: $26$) different structs to represent common classes of sets (such as half-spaces).
These set types simply expect and store the corresponding parameters that represent the set.
For example, the \code{HalfSpace} stores the vector \code{a} and the scalar \code{b}.
(There are a few exceptions where the constructor performs normalization by default, e.g., \code{HPolygon}, representing a two-dimensional polytope, sorts the constraints by the vectors \code{a} in counter-clockwise order.)
Hence construction is fast and the internal representation is space efficient.
For instance, the \code{BallInf} represents a hypercube specified by the center vector $c \in \R^n$ and the radius $r \in \R$.
In $n$ dimensions, a hypercube has $2^n$ vertices, but creating an $1{,}000$-dimensional \code{BallInf} is instantaneous.

\begin{minipage}{\linewidth}
\vspace{-\abovedisplayskip}
\begin{lstlisting}
julia> @time BallInf(zeros(1000), 1.0)
0.000005 seconds (2 allocations: 7.969 KiB)
\end{lstlisting}
\end{minipage}

\subsection{Extracting information from sets}

Being able to represent sets is not useful by itself because we also want to interact with them.
For example, we may want to draw samples from a set.
A general approach to do that is rejection sampling, which picks a random point $x \in \R^n$ and checks whether $x \in \X$ holds.
We can thus use rejection sampling with any set type that implements a membership test.

\begin{minipage}{\linewidth}
\vspace{-\abovedisplayskip}
\begin{lstlisting}
julia> ones(1000) ∈ BallInf(zeros(1000), 1.0)
true
\end{lstlisting}
\end{minipage}

Other typical properties that can be checked for sets $\X$ and $\Y$ are emptiness ($\X = \emptyset$; \code{isempty}), inclusion ($\X \subseteq \Y$; \code{issubset}), and having no point in common ($\X \cap \Y = \emptyset$; \code{isdisjoint}).
The complexity of such operations depends on the representation of the set.
Take for example a polyhedron represented by the list of its linear constraints.
Emptiness can be checked by formulating a feasibility linear program.
Inclusion and disjointness can be checked using the support function (see Section~\ref{sec:supfun}).

\smallskip

We may also want to obtain information that is encoded in the set representation.
For example, we can ask for the list of vertices of a polytope.
We have seen that a hypercube is represented by the center and the radius, so the vertices need to be computed on demand.

\begin{minipage}{\linewidth}
\vspace{-\abovedisplayskip}
\begin{lstlisting}
julia> vertices_list(BallInf([1.0, 4.0], 1.0))
4-element Vector{Vector{Float64}}:
 [2.0, 5.0]
 [0.0, 5.0]
 [2.0, 3.0]
 [0.0, 3.0]
\end{lstlisting}
\end{minipage}

Equality of sets in the mathematical sense can be checked via \code{isequivalent} (which by default checks mutual inclusion):

\begin{minipage}{\linewidth}
\vspace{-\abovedisplayskip}
\begin{lstlisting}
julia> X = Interval(-1, 1) × Interval(-1, 1)
CartesianProduct{Float64,
  Interval{...}, Interval{...}}(...)

julia> Y = BallInf(zeros(2), 1.0)

julia> isequivalent(X, Y)
true
\end{lstlisting}
\end{minipage}

\subsection{Set interfaces}

Sometimes the same implementation works for several set types.
LazySets uses a hierarchy of abstract types (which we call \emph{interfaces}) to summarize common functionalities.
For example, \code{AbstractHyperrectangle} is a supertype of all hyperrectangular set types such as \code{BallInf} and provides a default implementation to compute the volume.
When adding a new set type representing a hyperrectangle, it will automatically use this implementation.

The following list is not exhaustive, but should help as a mental model of how the library is organized. Definitions are given from more specific to more general (i.e., less structured).

\smallskip

\code{AbstractHyperrectangle}: Hyperrectangular sets can be represented by a center vector $c \in \R^n$ and a radius vector $r \in \R^n$. Each $x \in \X$ can be written as $x_i = c_i + \xi_i r_i$ for $i = 1,\ldots, n$, for some $\xi_i \in [-1, 1]$. Implementations include intervals (\code{Interval}), hypercubes (\code{BallInf}), and the general \code{Hyperrectangle}.

\smallskip

\code{AbstractZonotope}: Zonotopic sets are those which admit a representation given by a center $c \in \R^n$ and a finite set of \emph{generators} $g_j \in \R^n$, $j \in 1, \ldots, m$, such that $x \in \X$ is can be written as $x = c + \sum_j \xi_j g_j$ for some $\xi_j \in [-1, 1]$. Hyperrectangular sets are also zonotopic, as well as general zonotopes (\code{Zonotope}).

\smallskip

\code{AbstractPolyhedron}: A set is called polyhedral if it can be expressed as a finite intersection of half-spaces. Special cases include hyperrectangular and zonotopic sets, as well as more general polytopes (\code{HPolytope}, \code{VPolytope}) and also possibly unbounded polyhedra (\code{HPolyhedron}).

\smallskip

\code{LazySet}: All convex set types belong to this abstract supertype to prevent type piracy when extending \code{Base} functions.
We are working toward having non-convex sets, such as set unions, in the same type hierarchy as well.

\subsection{Set operations}

We have seen that we can interact with sets by checking properties.
Importantly, we can also apply set operations to sets for constructing new sets.
(By default the result is a new set instance and the original set instance is not manipulated.)
For details about the complexity for common set representations we refer to \cite[Table~1]{althoff2020set}.
For example, one common set operation is to translate (or shift) every element in the set by a constant vector.

\begin{minipage}{\linewidth}
\vspace{-\abovedisplayskip}
\begin{lstlisting}
julia> B1 = BallInf([1.5, 2.0], 1.0)
julia> B2 = translate(B1, [1.5, -1.0])
julia> dump(B2)
BallInf{Float64, Vector{Float64}}
  center: Array{Float64}((2,)) [3.0, 1.0]
  radius: Float64 1.0
\end{lstlisting}
\end{minipage}

As seen above, a translation usually preserves the set type.
For most operations this is generally not the case.
For instance, the intersection of two half-spaces is itself not a half-space but a polyhedron.

\begin{minipage}{\linewidth}
\vspace{-\abovedisplayskip}
\begin{lstlisting}
# intersect {x | x <= 1} and {x | x >= 0}
julia> P = intersection(HalfSpace([1.0], 1.0),
                        HalfSpace([-1.0], 0.0))
julia> typeof(P)
HPolyhedron{Float64, Vector{Float64}}
\end{lstlisting}
\end{minipage}

For a complete list of the set operations available in LazySets we refer to Table~\ref{tab:operations}.

\section{The LazySets paradigm} \label{sec:lazy}

Having described how to operate with basic sets, we consider a more fundamental representation problem.
We have seen that there exist classes of sets.
Some of them, such as polyhedra, are closed under various set operations. That property is convenient because the type of the result is known in advance.
The whole class of convex sets is also closed under the operations described in the upper part of Table~\ref{tab:operations}, but we can typically not represent the result with the limited amount of set types in LazySets. This is not a shortcoming of LazySets: you would need infinitely many set representations for all possible combinations! However, we can resort to a simple yet powerful trick to effectively represent the result of the set operations: \emph{lazy representation}.
Going a step further, by making both basic set types and operations between sets live in the same abstraction layer (namely always subtyping \code{LazySet} irrespective whether it is a concrete set or the result of an operation) allows to easily \emph{compose} set computations.

\subsection{Type composition} \label{sec:composition}

To give an example, the Minkowski sum of a square and a disc (two-dimensional balls in the infinity norm and Euclidean norm) is not representable with a basic type in LazySets. Hence \code{minkowski\_sum} will yield an error. But we can apply the \emph{lazy} operation, which is called \code{MinkowskiSum}. \code{MinkowskiSum} itself subtypes \code{LazySet}. This choice allows for ease of composition.

We wrap the operands in a new object that, by definition, represents the result of the operation, but without actually performing the computation. (This also motivates the name of LazySets).

\smallskip

As an illustrative example, suppose that we are interested in the formula $\Omega_0 = \CH(\X_0, \Phi \X_0 \oplus Y)$, where $\CH$ and $\oplus$ are defined in Table~\ref{tab:operations}, $X_0$ and $Y$ are sets, and $\Phi$ is a matrix defined in Appendix~\ref{sec:omega0}. Such formulas are prevalent in reachability analysis of linear initial-value problems, or nonlinear ones after some form of conservative linearization; see for example \cite{althoff2020set,ForetsS21} and references therein. Given the sets $\X_0$, $Y$, and the matrix $\Phi$, we can write:

\begin{minipage}{\linewidth}
\begin{lstlisting}
julia> Ω₀ = CH(X₀, Φ*X₀ ⊕ Y)
\end{lstlisting}
\end{minipage}

\begin{minipage}{0.69\linewidth}
Then, $\Omega_0$ is a (nested) \emph{lazy} representation of the  operation just as a normal \code{LazySet}. As such, it can be used for further operations (conversion, approximation, evaluation). The structure of the nested operations is internally represented in the form of a tree, which can be visualized with \href{https://github.com/JuliaTeX/TreeView.jl}{TreeView.jl} as shown in the diagram (right).
\end{minipage}
\hfill
\begin{minipage}{0.2\linewidth}
	\begin{tikzpicture}[level/.style={sibling distance=8mm,level distance=8mm}]
		\node {\texttt{CH}}
		child {
			node {\texttt{X}}
		}
		child {
			node {$\oplus$}
			child {
				node {\texttt{*}}
				child {
					node {$\Phi$}
				}
				child {
					node {\texttt{X}}
				}
			}
			child {
				node {\texttt{Y}}
			}
		};
	\end{tikzpicture}
\end{minipage}

\smallskip

Lazy operations can be efficiently evaluated, as we describe next.

\subsection{Support-function calculus} \label{sec:supfun}

A standard approach to operate with compact and convex sets $\X \in \R^n$ is to use the \emph{support function} \cite{LeGuernic09}.
The support function along direction $d \in \R^n$, denoted $\rho(d, \X)$, is the maximum of $d^T x$ over all $x \in \X$, i.e., the support function describes the (signed) distance of the supporting hyperplane in direction $d$ from the origin.
It can be used to efficiently find the boundary of a set in a given direction.
The maximizers are called \emph{support vectors} $\sigma(d, \X)$. Intuitively, the support vectors are the extreme points of $\X$ in direction $d$.
Basic properties of the support function are given in Appendix~\ref{sec:supfunc_properties}.

For various set representations, the support function is known analytically and can be efficiently evaluated numerically. Such cases include hyperrectangular sets and zonotopic sets.
For sets with less structure, e.g., if $\X$ is a polytope in half-space representation, its support function can be computed by solving a linear program, for which fast and robust solvers exist.
But the main advantage of using the support function in LazySets lies in the extensive use of \emph{composition rules}, as we describe later.

\smallskip

\begin{figure}
	\centering
	\includegraphics[width=0.49\linewidth, keepaspectratio]{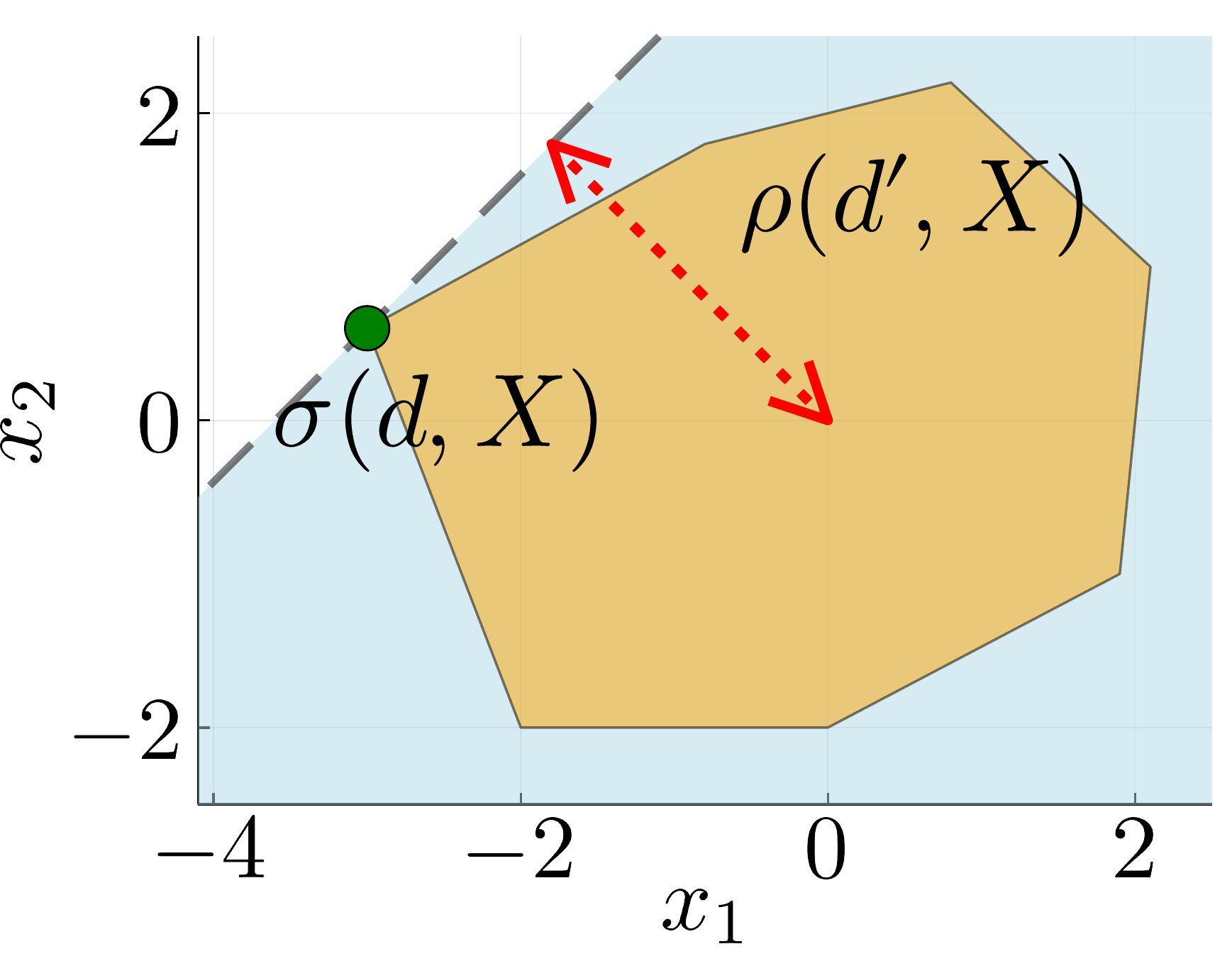}
	\includegraphics[width=0.49\linewidth, keepaspectratio]{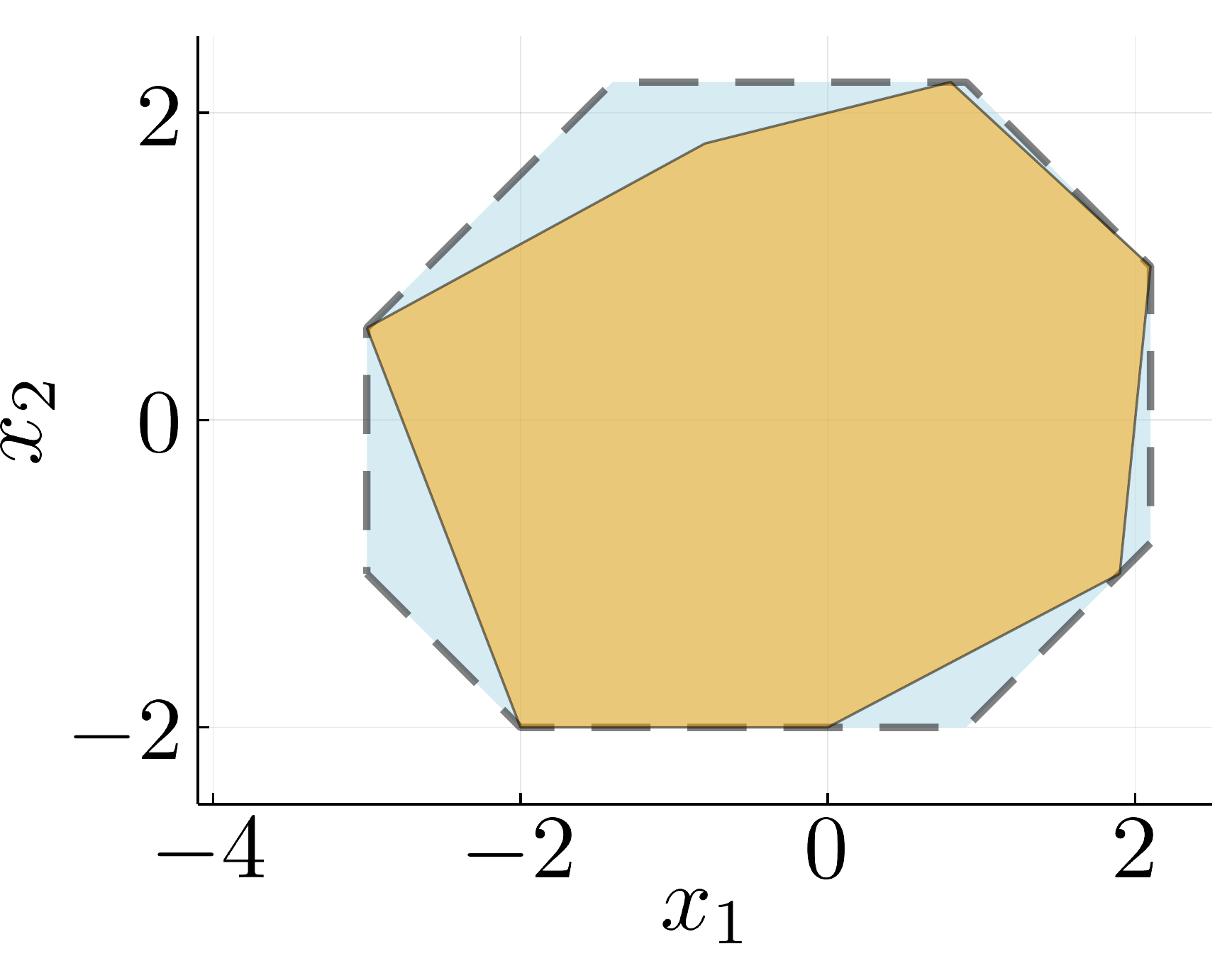}
	\vspace*{1mm}
	\caption{Left: The supporting hyperplane of the set $\X$ along direction $d$.
	In red we plot the distance of the hyperplane to the origin, which is given by $\rho(d', \X)$ where $d' = d / \Vert d \Vert$.
	Right: An outer approximation of $\X$ using the eight directions of a regular octagon.}
	\label{fig:supfunc}
\end{figure}

LazySets offers \code{$\uprho$(d, X)} (or \code{support\_function(d, X)}) to compute the support function $\rho(d,\X)$, and \code{$\upsigma$(d, X)} (or \code{support\_vector(d, X)}) to compute (some) support vector $\sigma(d,\X)$. Fig.~\ref{fig:supfunc} (left) illustrates the evaluation of the support function over the polygon $\X$ (orange) along direction $(-1, 1)^T$.

\begin{minipage}{\linewidth}
\vspace{-\abovedisplayskip}
\begin{lstlisting}
julia> d = [-1, 1]

julia> ρ(d, X) # or support_function(d, X)
3.6

julia> σ(d, X) # or support_vector(d, X)
[-3.0, 0.6]
\end{lstlisting}
\end{minipage}

Consider again the set $\Omega_0$ from the previous section. Suppose that we are interested in the support value of $\Omega_0$ along a given direction $d \in \R^2$.
Since the support function distributes over the Minkowski sum, $\rho(d, \X \oplus \Y) = \rho(d, \X) + \rho(d, \Y)$ for any pair of sets $\X, \Y \subseteq \R^n$, and since it holds that $\rho(d, M \X) = \rho(M^T d, \X)$ for any matrix $M \in \R^{n \times n}$, we can propagate the computation through the operation tree until a concrete set is found, and in many cases, an analytic formula is available.
That is precisely what LazySets does, automatically, when we make queries to the lazy set such as asking for its support function along $d = (-1, 1)^T$ from before.

\begin{minipage}{\linewidth}
	\vspace{-\abovedisplayskip}
	\begin{lstlisting}
julia> @btime ρ($d, $Ω₀)
117.236 ns (2 allocations: 192 bytes)
-0.8

julia> @btime ρ($d, concretize($Ω₀))
  23.700 μs (203 allocations: 16.03 KiB)
-0.8
	\end{lstlisting}
\end{minipage}

In the first benchmark we evaluate the support function on the lazy set.\footnote{The Julia macro \code{@btime} provided by the \href{https://github.com/JuliaCI/BenchmarkTools.jl}{BenchmarkTools.jl} package evaluates the given command multiple times and returns the smallest record.}
In the second benchmark we first convert to a concrete set representation (in this case a polygon in vertex representation) and then evaluate the support function.
The first case is two orders of magnitude faster.
This exemplifies that set computations can be implemented efficiently using the support function, which becomes more prominent in higher dimensions.

We note that in the above computation we have only obtained a bound in direction $d$, not in other directions. For many applications it is sufficient to evaluate the support function in only a few directions. For example, in the helicopter model presented in Section~\ref{sec:reachability} we are only interested in the vertical velocity, so we only need to evaluate the support function twice to obtain the upper and lower bound in that dimension.
Another example is to enclose $\Omega_0$ with a bounded set, for which we can pick a list of \emph{template} directions (see Section~\ref{sec:approximation}).

\section{Conversion between set types} \label{sec:approx}

LazySets provides ways to convert one set representation to another.
If a conversion is not possible due to restrictions in the represented class of sets, LazySets provides ways to obtain approximations.
Turning to an approximate but simpler set representation is also interesting for answering questions efficiently that would otherwise be computationally expensive.

\subsection{Conversion}

LazySets extends Julia's \code{convert} function for converting between set representations. The first argument is the target type and the second argument is the source set.
Below are three mathematically equivalent representations of the interval $\X = [0, 1] \subseteq \R$:

\begin{minipage}{\linewidth}
	\vspace{-\abovedisplayskip}
	\begin{lstlisting}
julia> X = Interval(0, 1)
Interval{Float64,
  IntervalArithmetic.Interval{Float64}}([0, 1])

julia> convert(Hyperrectangle, X)
Hyperrectangle{Float64, Vector{Float64},
  Vector{Float64}}([0.5], [0.5])

julia> convert(Zonotope, X)
Zonotope{Float64, Vector{Float64},
  Matrix{Float64}}([0.5], [0.5])
	\end{lstlisting}
\end{minipage}
There are even more possibilities, such as representing $\X$ as an intersection of half-spaces (try \code{convert(HPolytope, X)}).

\smallskip

With multiple dispatch it is easy to define less obvious conversions, e.g., to convert the Cartesian product of an interval and a two-dimensional hyperrectangle to a three-dimensional zonotope:

\begin{minipage}{\linewidth}
	\vspace{-\abovedisplayskip}
	\begin{lstlisting}
julia> X = rand(Interval)

julia> Y = rand(Hyperrectangle, dim=2)

julia> Z = convert(Zonotope, X × Y)
Zonotope{Float64, ...}

julia> dim(Z)
3
\end{lstlisting}
\end{minipage}

\subsection{Approximation}\label{sec:approximation}

\begin{figure}
	\hfill
	\includegraphics[height=30mm]{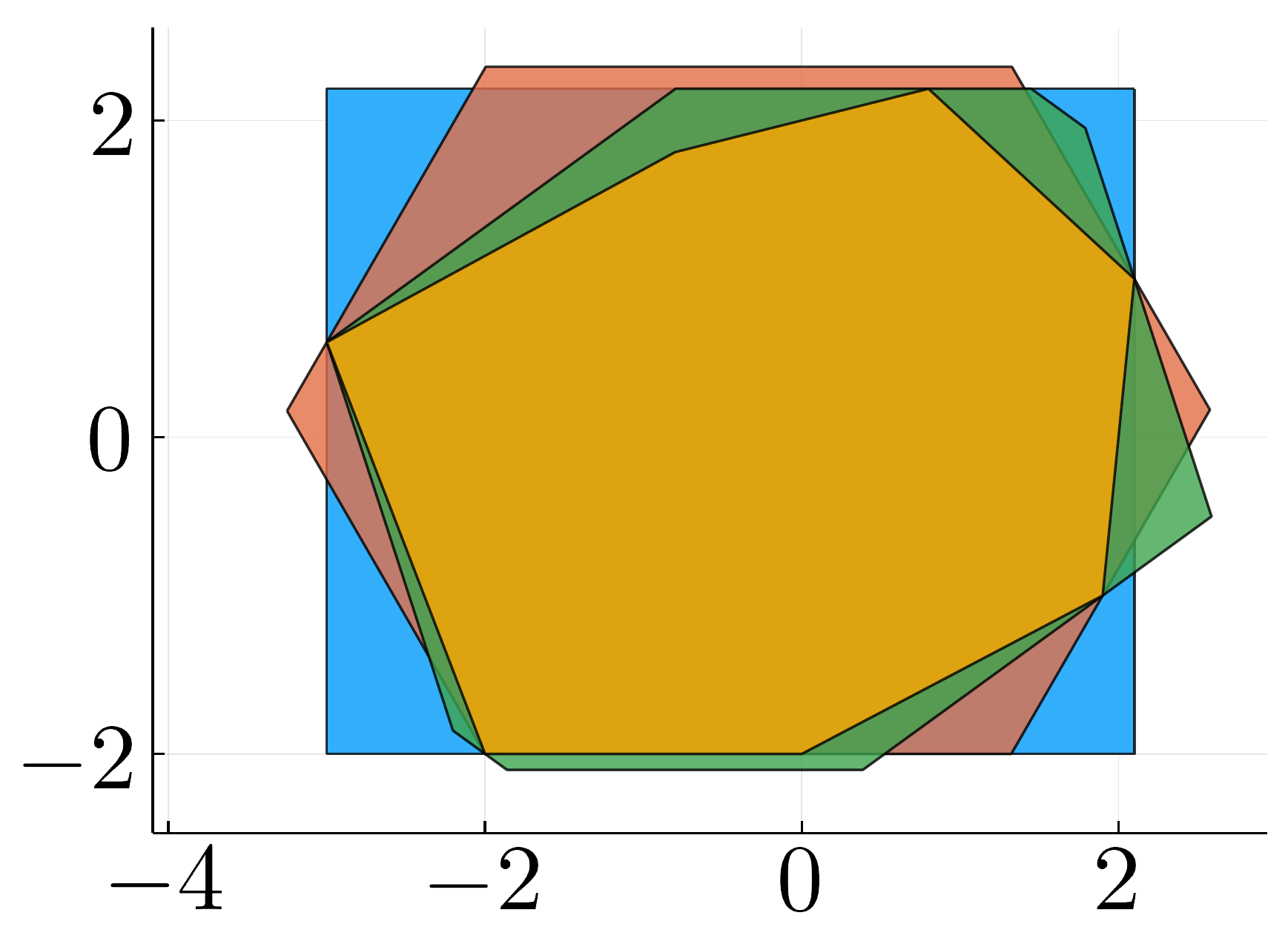}
	\hfill
	\includegraphics[height=30mm]{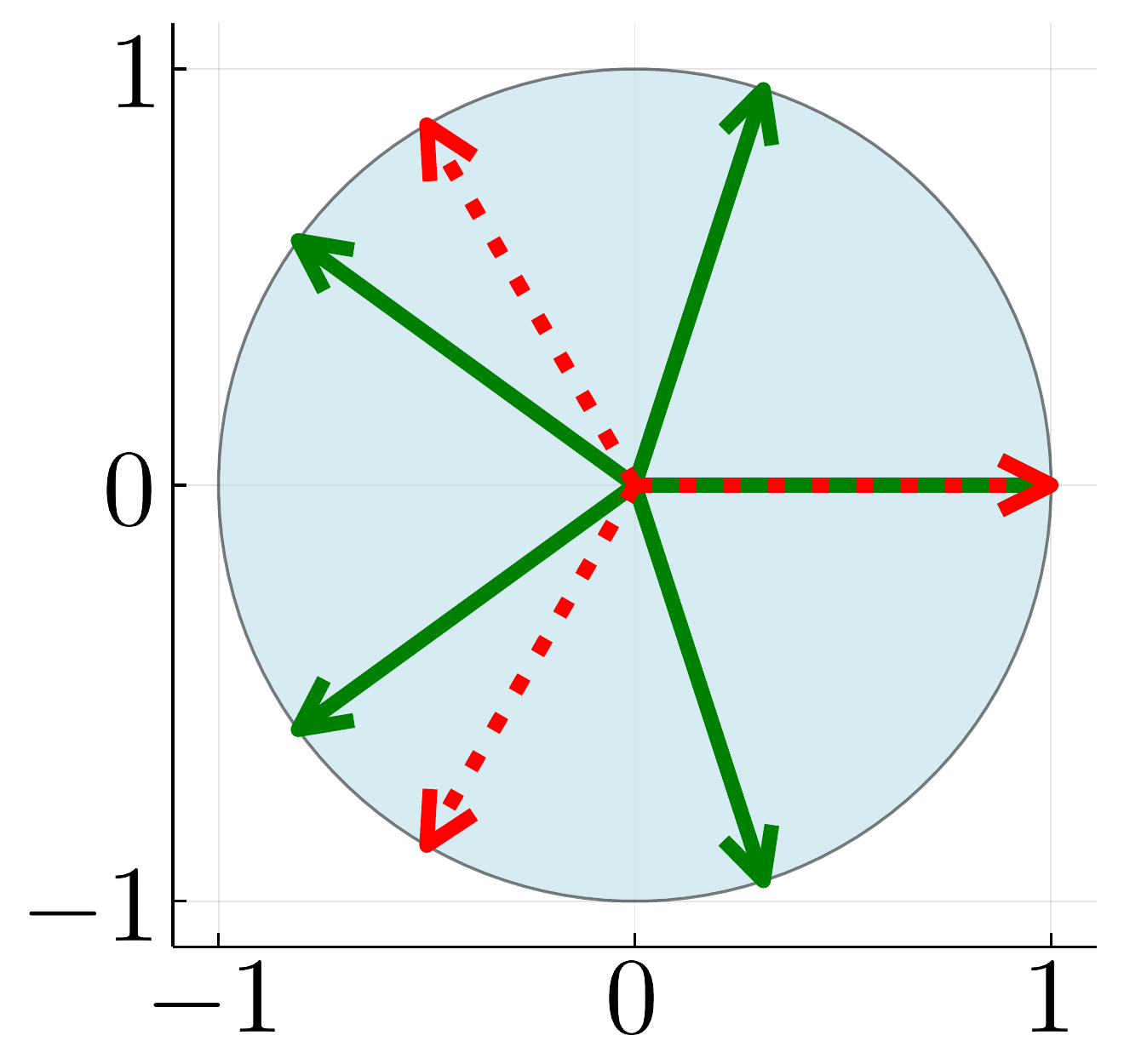}
	\hfill\
	\vspace*{1mm}
	\caption{Left picture: Overapproximation of the polytope from Fig.~\ref{fig:supfunc} (orange) with a hyperrectangle (blue) and two zonotopes. The zonotope generators were synthesized from three (red) resp.\ five (green) polar directions (right). Observe that the approximations are pairwise incomparable.}
	\label{fig:overapproximate}
\end{figure}

In many applications we do not require exact results but are content with an approximation. To still give mathematical guarantees, one usually aims for either over- or underapproximations.

We can use the support function to get an overapproximation: For every nonempty compact convex set $X \subseteq \R^n$ and $D \subseteq \R^n$ we have
\begin{equation*}
	X \subseteq \bigcap_{d \in D} \{d^T x \leq \rho(d, X)\}
\end{equation*}
and equality holds for $D = \R^n$.

\smallskip

LazySets has predefined common template directions such as \code{OctDirections(2)} for directions normal to a regular octagon in two dimensions. Fig.~\ref{fig:supfunc} (right) illustrates the evaluation of overapproximating the set $X$ with octagonal directions, resulting in a polygon with eight constraints. Apart from common fixed template directions there are also options for parametric uniform directions in two (\code{PolarDirections}) or three (\code{SphericalDirections}) dimensions or for a custom set of directions (\code{CustomDirections}).

\begin{minipage}{\linewidth}
\vspace{-\abovedisplayskip}
\begin{lstlisting}
julia> Xoct = overapproximate(X, OctDirections(2))

julia> length(constraints_list(Xoct))
8
\end{lstlisting}
\end{minipage}

In two dimensions, LazySets can compute $\varepsilon$-close overapproximations using a method by Kamenev~\cite{kamenev1996algorithm} later refined in~\cite{lotov2008modified}.
It is used via \code{overapproximate(X, $\varepsilon$)}, where $\varepsilon$ is the specified tolerance.
The higher-dimensional extension is not implemented yet.
On the other hand, a higher-dimensional set can be lazily projected using the support function to a lower-dimensional subspace, where the available method applies.

\smallskip

In some applications, we may want to ensure that the result has a specific set type.
The smallest bounding box is available by specifying the second argument type. It yields a \code{Hyperrectangle}, which is more efficient to work with.

\begin{minipage}{\linewidth}
	\vspace{-\abovedisplayskip}
	\begin{lstlisting}
julia> overapproximate(X, Hyperrectangle)

julia> box_approximation(X) # alias
\end{lstlisting}
\end{minipage}

We can use \code{overapproximate(P, Zonotope, D)}, where \code{P} is a polytope and \code{D} is a vector of directions used as candidates for the generators, to obtain a zonotope (in any dimension). We show some example overapproximations in Fig.~\ref{fig:overapproximate}.
Underapproximations can be obtained using the function \code{underapproximate}.

\section{Ecosystem} \label{sec:applications}

The Julia language features a rich ecosystem for scientific computing.
LazySets uses \href{https://github.com/JuliaPackaging/Requires.jl}{Requires.jl} to add further functionality depending on whether some external packages are loaded.
Here we quickly list various examples, but the list is not exhaustive.
From three-dimensional visualizations to reachability analysis using non-convex set representations, there is a broad spectrum of features that are available loading optional dependencies.

\smallskip

Some LazySets functionality prints instructive error messages when the corresponding packages are not loaded but are required by the code. For example, the conversion between constraint and vertex representation of general polytopes in dimension higher than two requires the optional package \href{https://github.com/JuliaPolyhedra/Polyhedra.jl}{Polyhedra.jl}:

\begin{minipage}{\linewidth}
\vspace{-\abovedisplayskip}
\begin{lstlisting}
julia> X = rand(VPolytope, dim=4)
	
julia> constraints_list(X)
ERROR: AssertionError: package 'Polyhedra' not
  loaded (it is required for executing
  `default_polyhedra_backend`)

# fix the error by loading the optional package
julia> import Polyhedra

julia> constraints_list(X)
12-element Vector{HalfSpace{Float64, ...}}}:
...
	\end{lstlisting}
\end{minipage}

\subsection{Plotting three-dimensional sets and projections}

While writing this article, we received the following question:

\begin{quote}
We need to plot polyhedra given in a form like this:

\begin{center}
	\texttt{2 * x1 >= 0 \& 3 * x2 + 1.7 * x3 >= 0}
\end{center}

Is there a way to plot a 2D projection with LazySets?
\end{quote}

Yes, there is!
We replied with the script below.
The script nicely illustrates the interaction of LazySets with the Julia ecosystem.
We use \href{https://github.com/JuliaSymbolics/Symbolics.jl}{Symbolics.jl} for reading the polyhedron in symbolic form, Polyhedra.jl and \href{https://github.com/JuliaPolyhedra/CDDLib.jl}{CDDLib.jl} for projecting the polyhedron, and \href{https://github.com/JuliaPlots/Plots.jl}{Plots.jl} to plot the two-dimensional projection.

\begin{minipage}{\linewidth}
\vspace{-\abovedisplayskip}
\begin{lstlisting}
using Symbolics
import Polyhedra, CDDLib, Plots

# projected polyhedron from symbolic constraints
vars = @variables x1, x2, x3  # create symbols
P = HPolyhedron([2*x1>=0, 3*x2+1.7*x3>=0], vars)
Q = project(P, [2, 3])  # 2D projection (x2 and x3)

Plots.plot(Q)  # plot the 2D projection
\end{lstlisting}
\end{minipage}

\begin{figure}
	\centering
	\hfill
	\includegraphics[width=0.49\linewidth, height=3cm, keepaspectratio,clip,trim=40mm 14mm 62mm 35mm]{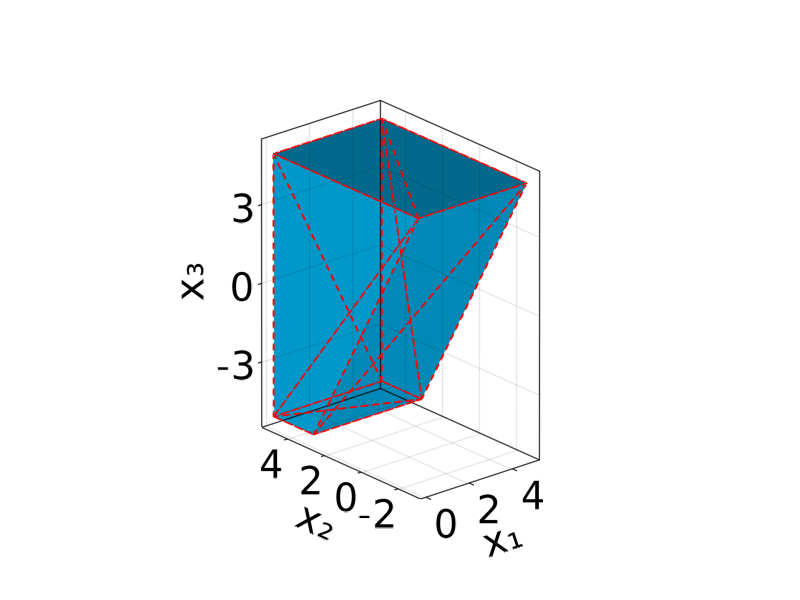}
	\hfill
	\includegraphics[width=0.49\linewidth, height=3cm, keepaspectratio]{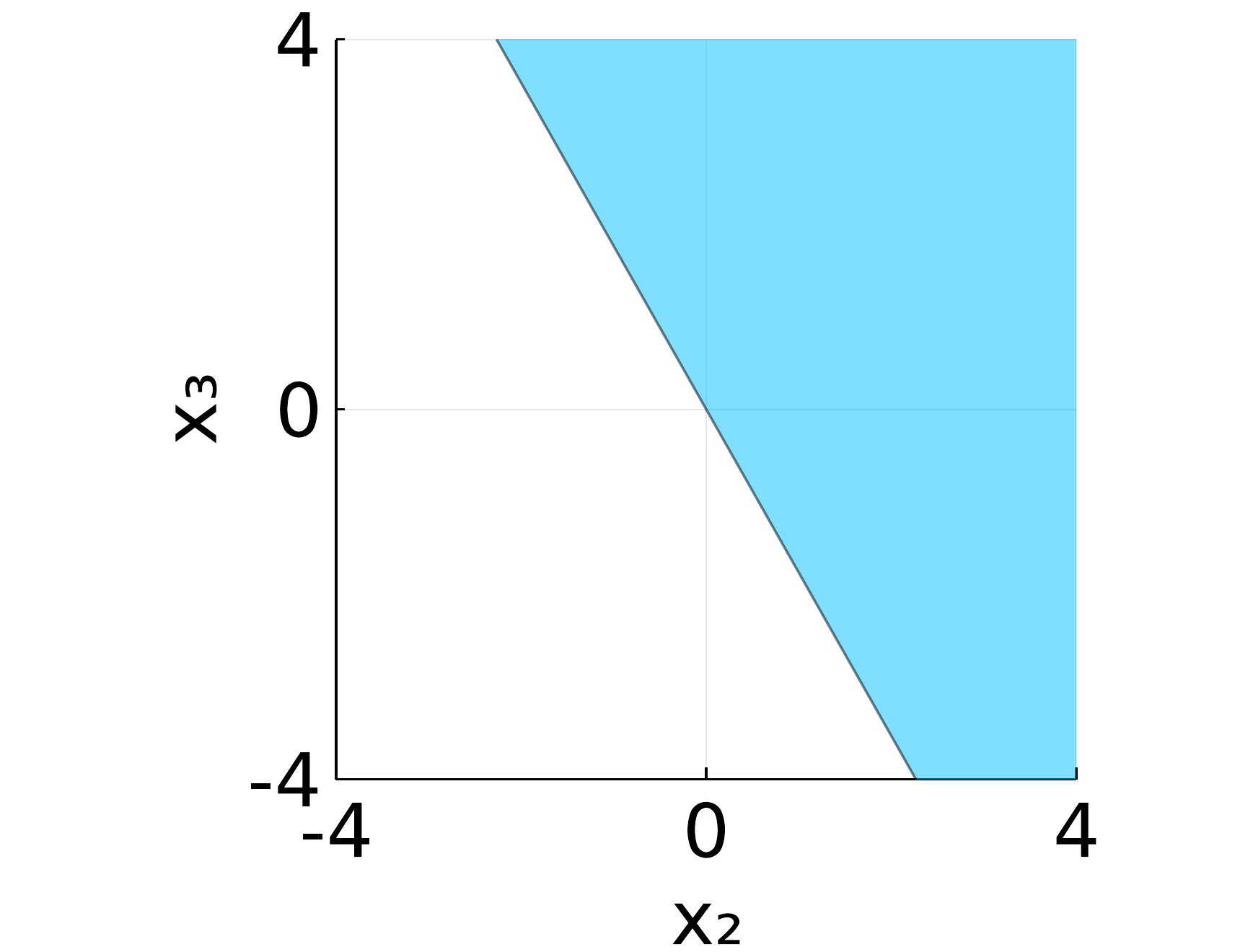}
	\hfill\,
	\vspace*{1mm}
	\caption{Three-dimensional polyhedron plotted using Makie (left) and a two-dimensional projection obtained by Polyhedra with CDDLib (right). Edges of a triangulation of the polyhedron are plotted in dashed lines (red).}
	\label{fig:polyhedra}
\end{figure}

We can also plot the three-dimensional set with \href{https://github.com/JuliaPlots/Makie.jl}{Makie.jl}.
The Makie plot and the two-dimensional projection are shown in Fig.~\ref{fig:polyhedra}.

\begin{minipage}{\linewidth}
\vspace{-\abovedisplayskip}
\begin{lstlisting}
import GLMakie

# intersection with a bounding box
B = BallInf(zeros(3), 5.0)
R = intersection(P, B)

plot3d(R)  # plot in three dimensions
\end{lstlisting}
\end{minipage}

\subsection{Generic numbers and automatic differentiation}\label{sec:numbertypes}

LazySets types are parametric in the number type. Hence it is simple to use custom number types.
Besides rationals and arbitrary-precision floating-point numbers, it is possible to make rigorous floating-point calculations using \href{https://github.com/JuliaIntervals/IntervalArithmetic.jl}{IntervalArithmetic.jl}.
Moreover, LazySets features a mechanism to globally tune the numeric tolerances used in floating-point operations.
To do so, use \code{set\_atol}, \code{set\_rtol}, and \code{set\_ztol} (for absolute, relative, and comparison-with-zero tolerance) respectively.
All set functions have been carefully designed to consistently use the specified tolerance and preserve it during operations.

\smallskip

Questions from users, bug reports, and feature requests are available in the \href{https://github.com/JuliaReach/LazySets.jl/issues/}{issue tracker}.
One question we received was from a user of \href{https://github.com/JuliaDiff/ForwardDiff.jl}{ForwardDiff.jl}, which is a package to compute the gradient:

\begin{quote}
	Using ForwardDiff.jl to get the gradient [...] throws the following error:

\texttt{ERROR: default tolerance for numeric type ForwardDiff.Dual\{...\} is not defined}
\end{quote}

The solution was surprisingly simple: extending the LazySets tolerance mechanism to work with dual numbers fixed the error.

\begin{minipage}{\linewidth}
	\vspace{-\abovedisplayskip}
	\begin{lstlisting}
import ForwardDiff
import LazySets.default_tolerance

default_tolerance(::Type{<:ForwardDiff.Dual}) = default_tolerance(Float64)
	\end{lstlisting}
\end{minipage}
With this code, the user could use automatic differentiation (AD) in the formula \code{area(intersection(X, Y))}, which computes the area of the intersection of sets $\X$ and $\Y$. More broadly, AD allows providing information to numerical methods around computational geometry, which is often used in robotics and machine learning~\cite{featherstone2014rigid,BaydinPRS17}.

\subsection{Taylor models as an example of non-convex sets}\label{sec:taylormodels}

LazySet is not limited to convex set representations. Apart from operating with set unions (\code{UnionSet}), which are generally non-convex, the library offers functionality to handle intrinsically non-convex set representations.
Examples include star sets (\code{AbstractStar}), polynomial zonotopes (\code{PolynomialZonotope}), and Taylor models. The latter representation is available in the package \href{https://github.com/JuliaIntervals/TaylorModels.jl}{TaylorModels.jl} \cite{TaylorModels.jl,BenetFSS19}, which LazySets interacts with.

\smallskip

\begin{figure}
	\centering
	\includegraphics[height=3.5cm, keepaspectratio]{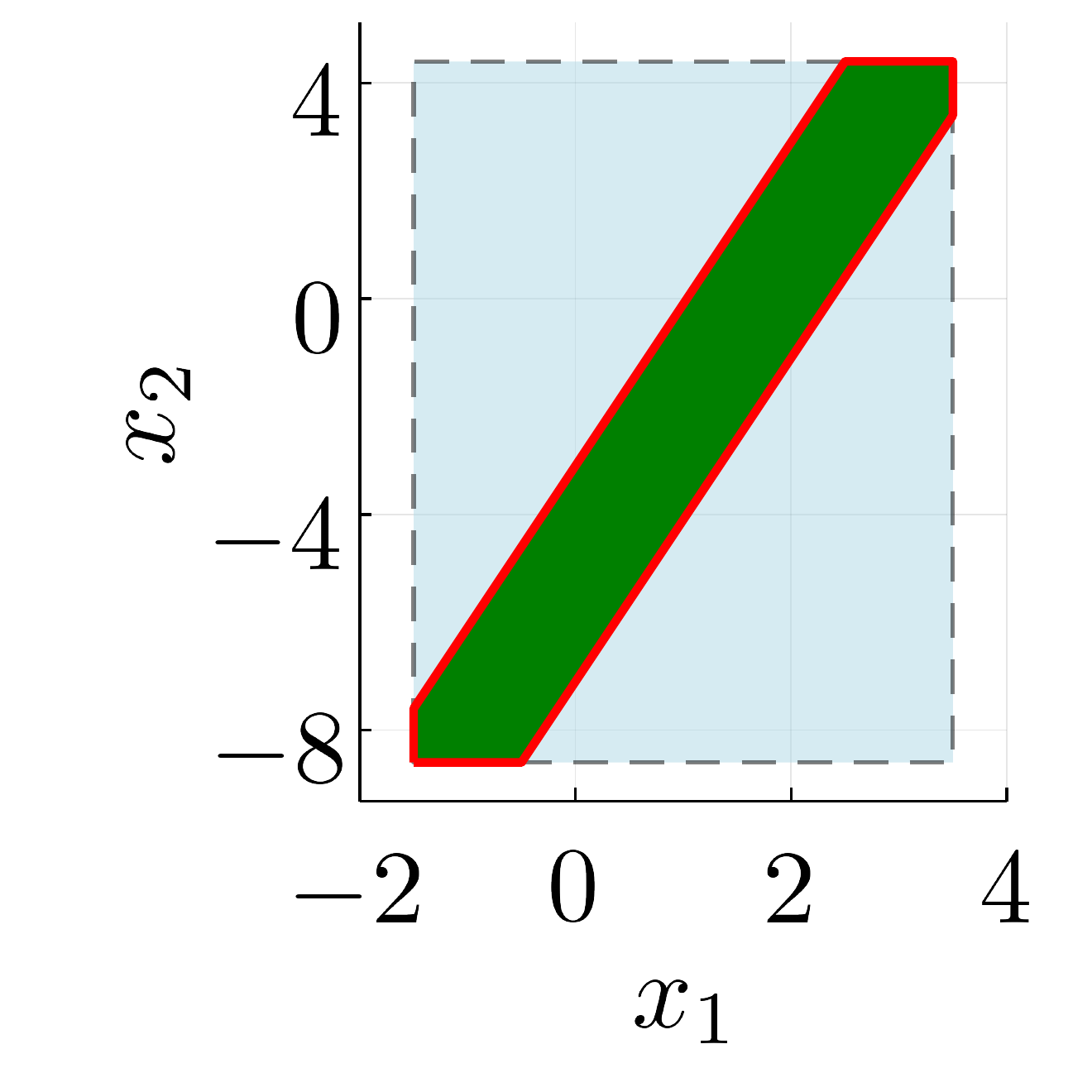}
	\includegraphics[height=3.5cm, keepaspectratio]{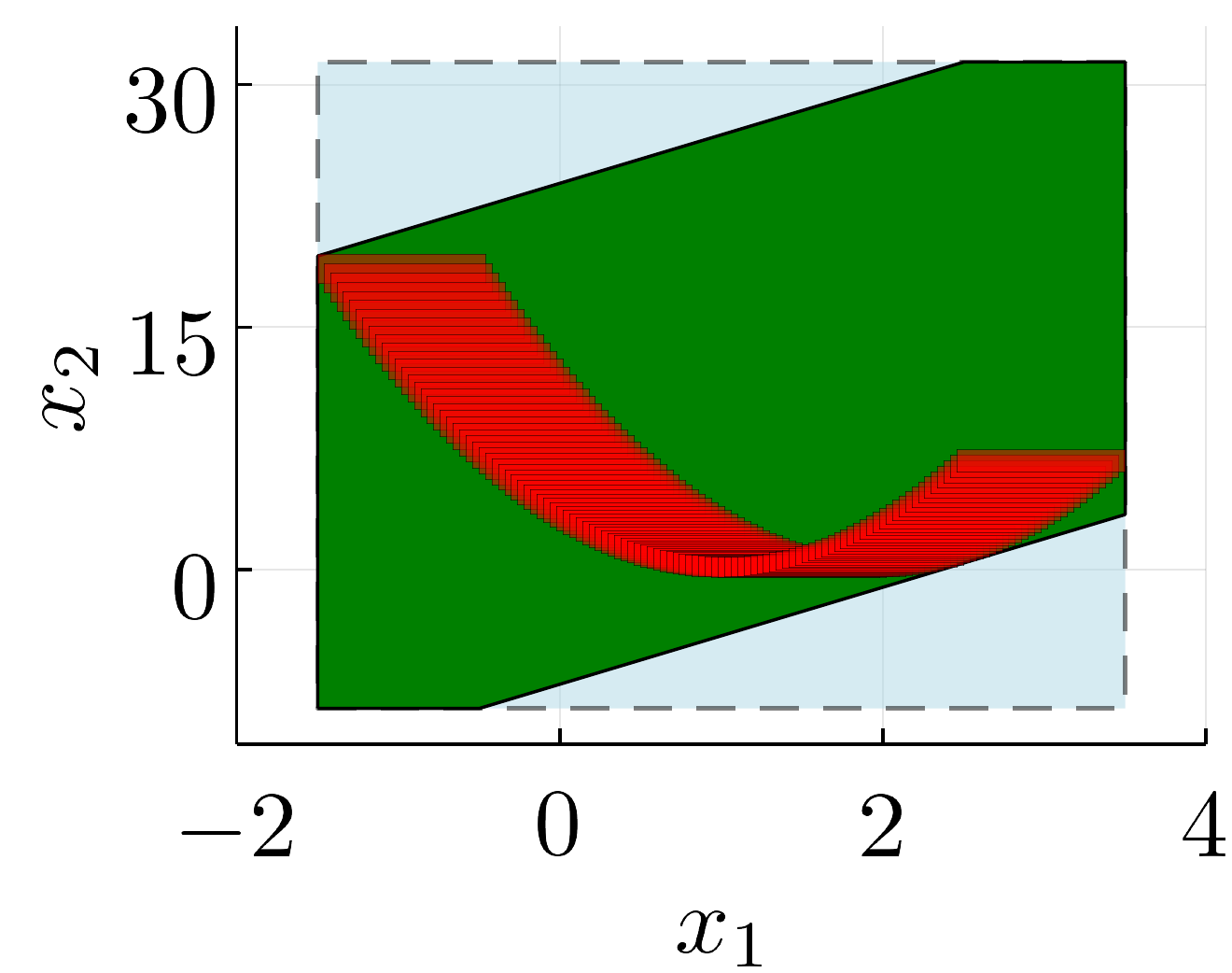}
	\caption{Left: Exact conversion of a linear Taylor model (red) to a zonotope $Z$ (green) against an inexact evaluation $H$ using interval arithmetic (blue, dashed). Right: Approximate conversion of a non-linear Taylor model (red) using a zonotope (green).}
	\label{fig:taylormodelsconversion}
\end{figure}

Informally, a Taylor model can be thought of as an interval tube around a polynomial.
See for example the red set in Fig.~\ref{fig:taylormodelsconversion}.
To illustrate, we use LazySets to convert, or evaluate the range, of a Taylor model (the conversion is formally explained in~\cite{schilling2021verification}).
It is easy to operate with Taylor models in LazySets (the definition of \code{vTM} for the linear and nonlinear case is given in Appendix~\ref{sec:taylormodels_appendix}):

\begin{minipage}{\linewidth}
	\vspace{-\abovedisplayskip}
	\begin{lstlisting}
using TaylorModels

# approximate a vector of Taylor models
# with a zonotope
Z = overapproximate(vTM, Zonotope)

# approximate a vector of Taylor models
# with a hyperrectangle
H = overapproximate(vTM, Hyperrectangle)
	\end{lstlisting}
\end{minipage}

In the case when the Taylor model is linear, the conversion is exact, because the zonotope stores linear dependencies between variables.
Hence, as is shown in Fig.~\ref{fig:taylormodelsconversion} (left), the zonotope approximation (\code{Z}) is better than directly using interval arithmetic to evaluate the range (\code{H}), and queries about the zonotope provide quick information about the exact set, more accurate than the box approximation.

%
%

\smallskip

If the Taylor model contains nonlinear terms, the zonotope provides only an enclosure but can still be more precise than the box approximation; a comparison is shown in Fig.~\ref{fig:taylormodelsconversion} (right); we obtain a more precise approximation of the Taylor model by partitioning its domain and evaluating each resulting block with a smaller box.

\smallskip

We also mention that LazySets can be combined with an interval-constraint-programming approach by simply loading the optional dependency \href{https://github.com/JuliaIntervals/IntervalConstraintProgramming.jl}{IntervalConstraintProgramming.jl}.

\subsection{Reachability applications}\label{sec:reachability}

LazySets is the core library of JuliaReach, a Julia ecosystem to perform reachability analysis of dynamical systems of the form $\dot{x}(t) = f(x(t), u(t))$.
JuliaReach builds on sound scientific approaches such as \cite{BogomolovFFVPS18} and was, in two occasions (2018 and 2020), the winner of the annual friendly competition on Applied Verification for Continuous and Hybrid Systems (\href{https://cps-vo.org/group/ARCH}{ARCH-COMP}).

\smallskip

\begin{figure}[tb]
	\centering
	\includegraphics[width=\linewidth,keepaspectratio]{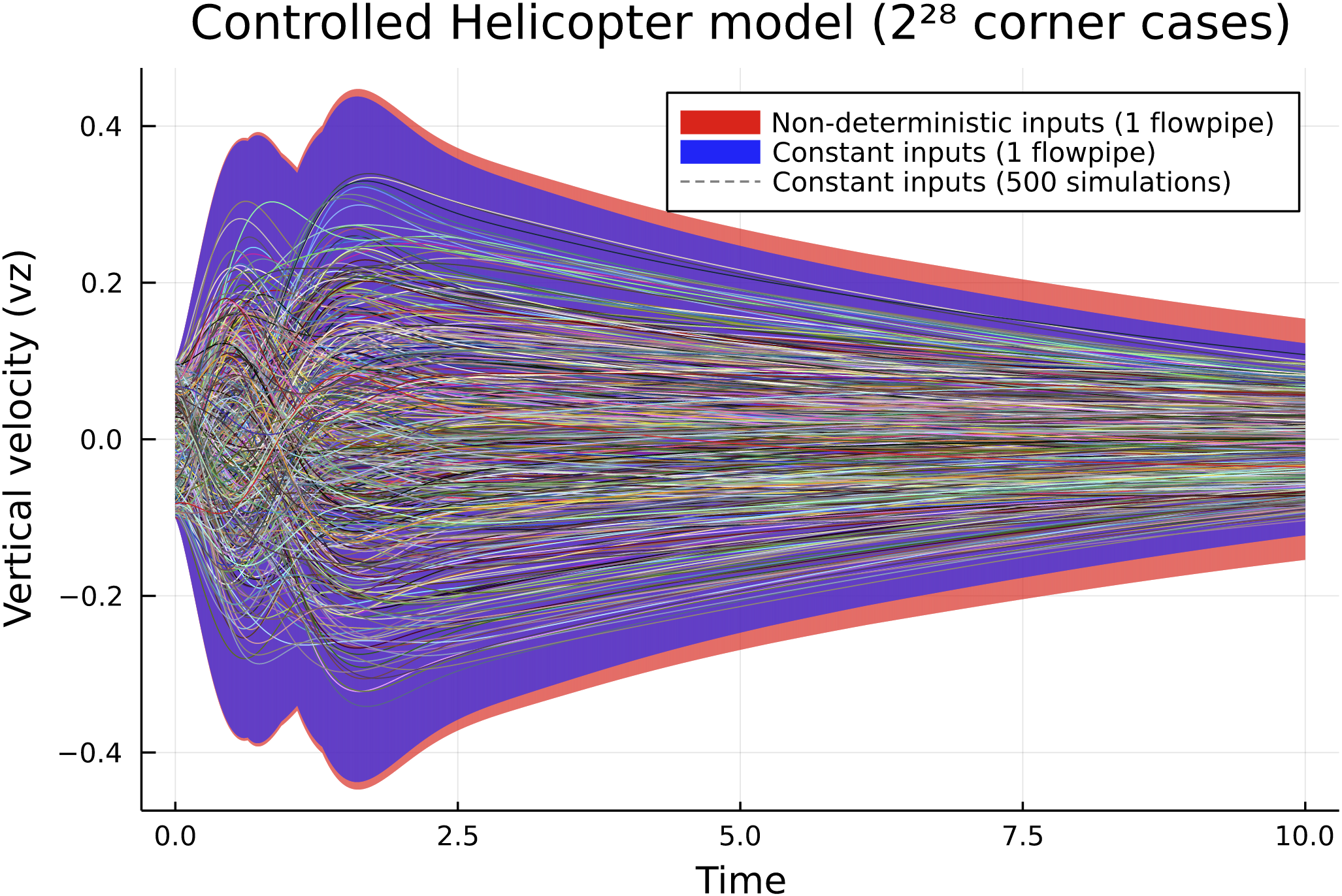}
	\vspace*{.5mm}
	\caption{Reachable states for the vertical velocity of the helicopter model.
	We consider two cases for the input signals $u(t)$: constant and non-deterministic. Five hundred trajectories drawn randomly from the set of initial states and with constant inputs are shown on top.}
	\label{fig:flowpipe_heli}
\end{figure}

Reachability techniques are implemented in the JuliaReach package \href{https://github.com/JuliaReach/ReachabilityAnalysis.jl}{ReachabilityAnalysis.jl}, which uses LazySets at its core for dealing with sets, including the computation of reachable states. The union of reachable states for consecutive time intervals is often called a \emph{flowpipe}.
We include two examples. In Fig.~\ref{fig:flowpipe_heli}, a flowpipe for the vertical velocity of an 8-dimensional helicopter model with a 20-dimensional controller from \cite{skogestad2007multivariable} is shown.
The model has an uncertain initial condition $x(0)$ originating from a set $\X_0$, as well as a non-deterministic input signal $u(t) \in \U$ (i.e., the input signal $u(t)$ can vary arbitrarily within the bounds specified by the set $\U$).
The computation terminates in nearly $50$\,ms, illustrating the precision and speed of LazySets.
It should be noted that the set of initial states has $2^{28}$ corner cases, thus even in the simpler setting where the inputs are held constant, exhaustive evaluation using simulations is computationally intractable, since it would require $268$ million runs.

\smallskip

\begin{figure}[tb]
	\centering
	\includegraphics[width=\linewidth,keepaspectratio,height=4.5cm]{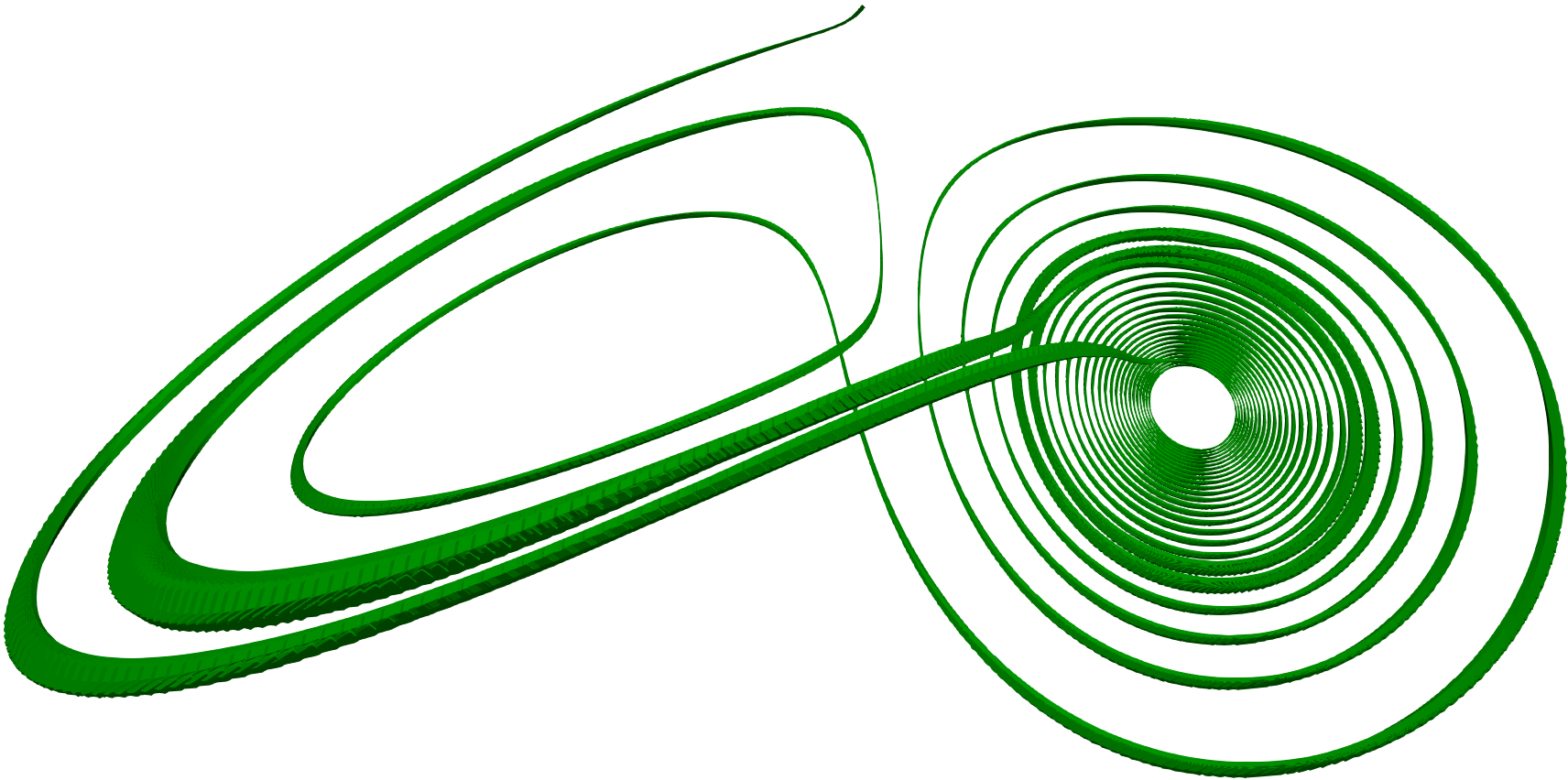}
	\vspace*{.5mm}
	\caption{Zonotope overapproximation of a Taylor-model flowpipe for the Lorenz equations. The plot is rendered with Paraview.}
	\label{fig:flowpipe_lorenz}
\end{figure}

As a second example, a three-dimensional flowpipe for the well-known Lorenz system is shown in Fig.~\ref{fig:flowpipe_lorenz}.
The initial condition is taken from a flat hyperrectangle of radius $0.1$ along the $x$ coordinate.
%
%
The reachable states are represented using Taylor models, which are then approximated with zonotopes for further computations.
In this case we have exported the LazySets objects to a VTK file using the \href{https://github.com/jipolanco/WriteVTK.jl}{WriteVTK.jl} optional dependency, and rendered the picture with the open-source visualization tool \href{https://www.paraview.org/}{Paraview}.

\subsection{Parametrization and custom array types} \label{sec:custom_arrays}

For set types that contain array fields, we use type parameters. Hence it is possible to instantiate LazySets types with any custom array. Typically, such special arrays (e.g. dense, sparse, static) are used in applications that require high performance.

\smallskip

For example, static arrays (where the size of the array can be determined from the type) are preferable for efficient set computations with ``small'' arrays and are available upon loading \href{https://github.com/JuliaArrays/StaticArrays.jl}{StaticArrays.jl}:

\begin{minipage}{\linewidth}
	\vspace{-\abovedisplayskip}
	\begin{lstlisting}
julia> using StaticArrays

# random ten-dimensional zonotope
julia> Z = rand(Zonotope, dim=10, num_generators=30)

# convert to static arrays
julia> Zs = Zonotope(SVector{10, Float64}(Z.center),
SMatrix{10, 30, Float64}(Z.generators))
julia> d = ones(10)
julia> ds = SVector{10, Float64}(d)

# using normal arrays
julia> @btime ρ($d, $Z)
145.290 ns (0 allocations: 0 bytes)
72.68047192978314

# using static arrays
julia> @btime ρ($ds, $Zs)
44.899 ns (0 allocations: 0 bytes)
72.68047192978314
	\end{lstlisting}
\end{minipage}

\smallskip

Custom arrays can also be used to express knowledge about the \emph{structure} of the set, using only type information. For instance, an axis-aligned half-space can be defined such that its normal vector is a \code{LazySets.SingleEntryVector}, i.e., a vector with a single non-zero element.
Then, the concrete intersection with a hyperrectangular set can be computed very efficiently:

\begin{minipage}{\linewidth}
\vspace{-\abovedisplayskip}
\begin{lstlisting}
# half-space with a normal array
julia> @btime intersection($X, $Hvec)
419.424 μs (1799 allocations: 177.83 KiB)

# half-space with a specialized array
julia> @btime intersection($X, $Hsev)
376.059 ns (13 allocations: 1.08 KiB)
\end{lstlisting}
\end{minipage}
This example shows that a 100x speedup is obtained by exploiting structure encoded in the half-space's type.

\subsection{Interoperability with other languages}\label{sec:python}

It is not uncommon that scientists beginning to use Julia are familiar with other languages, specially with the Python programming language.
Below we show how to use \href{https://github.com/JuliaPy/pyjulia}{pyjulia} for working with LazySets types \emph{from Python}.
We can see that it is not necessary to use Julia objects everywhere; NumPy arrays can also be used, making the interoperability between Julia and Python effortless.

\begin{minipage}{\linewidth}
	\vspace{-\abovedisplayskip}
	\begin{lstlisting}[language=python]
$ python3 -m pip install --user julia

$ python3

>>> import julia
>>> julia.install()  # only once

>>> from julia import Base, LazySets
>>> from julia.LazySets import BallInf, volume

>>> B = BallInf(Base.zeros(3), 1.0)
>>> volume(B)
8.0

>>> import numpy as np
>>> c = np.array([0.0, 0.0, 0.0])
>>> B = BallInf(c, 1.0)
>>> volume(B)
8.0
	\end{lstlisting}
\end{minipage}

\section{Conclusion and Perspectives} \label{sec:conclusion}

We conclude with a brief discussion of the past, present, and future development perspectives of LazySets.

\subsection{Related libraries}

There are few publicly available libraries with a similar aim as LazySets.
All these libraries are used in the context of reachability analysis.
\href{https://github.com/hypro/hypro}{HyPro} is a C++ library for concrete representation and manipulation of sets such as convex polytopes and Taylor models and also offers a support-function representation of set operations \cite{SchuppAMK17}.
\href{https://github.com/TUMcps/CORA}{CORA} is an actively developed Matlab library centered around zonotopes and contains implementations of zonotope bundles, matrix zonotopes, and polynomial zonotopes~\cite{Althoff15}.
The \href{https://github.com/SystemAnalysisDpt-CMC-MSU/ellipsoids}{ellipsoidal toolbox} is a Matlab library for ellipsoids \cite{kurzhanskiy2006ellipsoidal}.
\href{http://spaceex.imag.fr/}{SpaceEx} is a C++ reachability library that established the use of the support function in reachability analysis, but it is not open source \cite{frehse2011spaceex}.

\smallskip

Finally, we mention other related Julia packages with a different aim.
As mentioned in Section~\ref{sec:applications}, \href{https://github.com/JuliaPolyhedra/Polyhedra.jl}{Polyhedra.jl} \cite{legat2021polyhedra} provides an interface for polyhedral computations and the double-description method; hence it complements nicely the lazy features available in our library.
\href{https://github.com/JuliaGeometry/GeometryBasics.jl}{GeometryBasics.jl} offers standard geometry types, creating a basis for graphics/plotting in Julia.
The more recent package
\href{https://github.com/JuliaGeometry/Meshes.jl}{Meshes.jl} is specialized on efficient and pure-Julia implementations of computational geometry and meshing algorithms.
Another Julia package for describing domains in Euclidean space is \href{https://github.com/JuliaApproximation/DomainSets.jl}{DomainSets.jl}; this package is being adopted for modeling PDE (partial differential equation) domains with \href{https://github.com/SciML/ModelingToolkit.jl}{ModelingToolkit.jl}.
Optimization is another scientific field where sets play a major role, with great contributions from Julia developers.
\href{https://github.com/jump-dev/MathOptInterface.jl}{MathOptInterface.jl}
\cite{MathOptInterface-2021} is at the core of \href{https://github.com/jump-dev/JuMP.jl}{JuMP.jl}~\cite{DunningHuchetteLubin2017}, Julia's mainstream modeling language for mathematical optimization.
Applications include set programming in \href{https://github.com/blegat/SetProg.jl}{SetProg.jl}
and set distances in \href{https://github.com/jump-dev/MathOptSetDistances.jl}{MathOptSetDistances.jl}.

\subsection{Origin of LazySets and current applications}

LazySets has its origins in a tool for reachability analysis of linear dynamical systems, using a compositional approach based on reducing high-dimensional lazy set representations into a sequence of low-dimensional projections that can be computed efficiently~\cite{BogomolovFFVPS18}.
This method presented the first approach to formally verify a $10{,}000$-dimensional benchmark from control engineering.
The reachability tool has since been rewritten in ReachabilityAnalysis.jl.
A preliminary exposition of these tools appeared in \cite{BogomolovFFPS19}.

\smallskip

We have recently applied LazySets to compute reachable states for linear wave propagation problems and heat transfer problems~\cite{forets2021combining}.
The scalability of the approach relies on exploiting the structure of linear systems through the support-function calculus and lazy evaluation.
Moreover, linear systems can be embedded in algorithms to analyze nonlinear systems \cite{forets2021reachability}.
Further case studies and comparison with other state-of-the-art tools can be found in \cite{ARCHCOMP20LINEAR, ARCHCOMP20NONLINEAR}.
LazySets has also been applied to the challenging domain of hybrid systems (systems with mixed discrete-continuous dynamics) for set propagation \cite{bogomolov2020reachability} and synthesis \cite{soto2021synthesis}.
Such problems require switching between different set representations and handling intersections efficiently and accurately.
Timed systems with non-deterministic events have been considered in \cite{forets2020efficient}; the approach is able to handle a large number of sets ($100$ million sets in zonotope representation), and it was shown to be an order of magnitude faster than competing tools~\cite{ARCHCOMP20LINEAR} such as CORA and SpaceEx.

\smallskip

Besides reachability analysis, LazySets can be used for other purposes. The tool has been adopted in a review article in the context of propagating sets through neural networks \cite{LiuALSBK21}, and new tools use LazySets for verification, e.g.,
\href{https://github.com/JuliaReach/NeuralNetworkAnalysis.jl}{NeuralNetworkAnalysis.jl}~\cite{schilling2021verification} and \href{https://github.com/sisl/OVERTVerify.jl}{OVERTVerify.jl}~\cite{sidrane2021overt}.
Other Julia packages using LazySets functionality include computation of invariant sets in \href{https://github.com/ueliwechsler/InvariantSets.jl}{InvariantSets.jl}, ray tracing for geometric optics in \href{https://github.com/microsoft/OpticSim.jl}{OpticSim.jl}, astronomical photometry in \href{https://github.com/JuliaAstro/Photometry.jl}{Photometry.jl}, thin film simulations in \href{https://github.com/Zitzeronion/Swalbe.jl}{Swalbe.jl}, and linear algebra with interval matrices in \href{https://github.com/JuliaIntervals/IntervalLinearAlgebra.jl}{IntervalLinearAlgebra.jl}.

\subsection{LazySets in numbers}

At the time of writing (LazySets-v1.52.1), the package consists of 125 source files with almost 26k lines of code (LOC), 66 test files with over 5k LOC, and 66 documentation files (markdown) with over 4k LOC.
It is maintained by the authors of this article\footnote{\texttt{@mforets} and \texttt{@schillic} handles on \href{https://github.com/}{github.com}, respectively.}.
The project has received contributions from 13 other people.
LazySets was used in 19 research articles.

\subsection{Future work}

Set computations often do not allow for typical tricks you would expect to see in a Julia package. For instance, when working with generic polyhedra, there is very little structure, so most information cannot be statically inferred and needs to be computed from the concrete values (such as whether the polyhedron is empty). That is why there are so many set types: to bring in more structure for algorithms and dispatch.
Another challenge in set computations is to preserve type stability: in some cases, the output set type cannot be predicted in advance.

\smallskip

While many algorithms are already optimized, some functions still use a suboptimal, generic fallback.
We are interested to identify and fix such cases.
In our experience, one can often obtain speed-ups within several orders of magnitude by adding new methods for Julia's multiple dispatch, as we exemplified in Section~\ref{sec:custom_arrays} for the intersection of an axis-aligned half-space with a hyperrectangle.

\smallskip

Another direction is the use of trait-based dispatch, which may be useful as a workaround for limitations of the Julia type system, e.g., that it does not allow for multiple inheritance.
Expressing properties of sets that fall outside the established LazySets type hierarchy would allow for even further flexibility.

\smallskip

Our next aimed milestone is proper support of non-convex set representations. While functionality to operate with such set representations is already available, the interoperability between convex and non-convex sets has room for improvement.

\smallskip

Finally, LazySets has a solid documentation of its API by an extensive use of docstrings and uses \href{https://github.com/JuliaDocs/Documenter.jl}{Documenter.jl} for its online documentation.
We plan to add more introductory examples and tutorials for first-time users.

\section*{Acknowledgments}

First, we thank all those of have contributed to LazySets\footnote{The full list of contributors is available \href{https://github.com/JuliaReach/LazySets.jl/graphs/contributors}{on Github}.}.
We are specially thankful to the following people for discussions and contributions at various stages of this work (sorted alphabetically): Tomer Arnon, Luis Benet, Aadesh Deshmukh, Goran Frehse, Daniel Freire, Bruno Garate, Ander Gray, Sebastian Guadalupe, Nikolaos Kekatos, Beno\^it Legat, Jorge P\'{e}rez Zerpa, Kostiantyn Potomkin, David P. Sanders, Frederic Viry, Ueli Wechsler and Peng Yu.
Finally, we thank the people behind Julia Seasons of Contributions and Google Summer of Code programs for their financial support of several students to further develop LazySets.
The first author is partly supported by Agencia Nacional de Investigaci\'{o}n e Innovaci\'{o}n (ANII), Uruguay.
Finally, we greatly appreciate Beno\^it Legat and Guillaume Dalle for their work in revising this paper.

\bibliographystyle{juliacon}
\bibliography{ref.bib}

\begin{thebibliography}{10}

\bibitem{Althoff15}
Matthias Althoff.
An introduction to {CORA} 2015.
In {\em ARCH@CPSWeek}, volume~34 of {\em EPiC Series in Computing}, pages
  120--151. EasyChair, 2015.
\href{http://dx.doi.org/10.29007/zbkv}{doi:10.29007/zbkv}.

\bibitem{ARCHCOMP20LINEAR}
Matthias Althoff, Stanley Bak, Zongnan Bao, Marcelo Forets, Goran Frehse,
  Daniel Freire, Niklas Kochdumper, Yangge Li, Sayan Mitra, Rajarshi Ray,
  Christian Schilling, Stefan Schupp, and Mark Wetzlinger.
{ARCH-COMP20} category report: Continuous and hybrid systems with linear
  continuous dynamics.
In {\em {ARCH}}, volume~74 of {\em EPiC Series in Computing}, pages 16--48.
  EasyChair, 2020.
\href{http://dx.doi.org/10.29007/7dt2}{doi:10.29007/7dt2}.

\bibitem{althoff2020set}
Matthias Althoff, Goran Frehse, and Antoine Girard.
Set propagation techniques for reachability analysis.
{\em Annual Review of Control, Robotics, and Autonomous Systems},
  4(1):369--395, 2020.
\href{http://dx.doi.org/10.1146/annurev-control-071420-081941}{doi:10.1146/annurev-control-071420-081941}.

\bibitem{BaydinPRS17}
Atilim~Gunes Baydin, Barak~A. Pearlmutter, Alexey~Andreyevich Radul, and
  Jeffrey~Mark Siskind.
Automatic differentiation in machine learning: a survey.
{\em J. Mach. Learn. Res.}, 18:153:1--153:43, 2017.

\bibitem{BenetFSS19}
Luis Benet, Marcelo Forets, David~P. Sanders, and Christian Schilling.
{TaylorModels}.jl: {T}aylor models in {J}ulia and its application to validated
  solutions of {ODE}s.
\url{https://swim2019.ensta-paristech.fr/}, 2019.

\bibitem{TaylorModels.jl}
Luis Benet and David~P. Sanders.
{JuliaIntervals/{T}aylor{M}odels.jl}.
\url{https://github.com/JuliaIntervals/TaylorModels.jl}, 2021.
\href{http://dx.doi.org/10.5281/zenodo.2613102}{doi:10.5281/zenodo.2613102}.

\bibitem{bezanson2017julia}
Jeff Bezanson, Alan Edelman, Stefan Karpinski, and Viral~B Shah.
Julia: A fresh approach to numerical computing.
{\em SIAM review}, 59(1):65--98, 2017.
\href{http://dx.doi.org/10.1137/141000671}{doi:10.1137/141000671}.

\bibitem{BogomolovFFPS19}
Sergiy Bogomolov, Marcelo Forets, Goran Frehse, Kostiantyn Potomkin, and
  Christian Schilling.
{JuliaReach}: a toolbox for set-based reachability.
In {\em {HSCC}}, pages 39--44. {ACM}, 2019.
\href{http://dx.doi.org/10.1145/3302504.3311804}{doi:10.1145/3302504.3311804}.

\bibitem{bogomolov2020reachability}
Sergiy Bogomolov, Marcelo Forets, Goran Frehse, Kostiantyn Potomkin, and
  Christian Schilling.
Reachability analysis of linear hybrid systems via block decomposition.
{\em {IEEE} Trans. Comput. Aided Des. Integr. Circuits Syst.},
  39(11):4018--4029, 2020.
\href{http://dx.doi.org/10.1109/TCAD.2020.3012859}{doi:10.1109/TCAD.2020.3012859}.

\bibitem{BogomolovFFVPS18}
Sergiy Bogomolov, Marcelo Forets, Goran Frehse, Fr{\'{e}}d{\'{e}}ric Viry,
  Andreas Podelski, and Christian Schilling.
Reach set approximation through decomposition with low-dimensional sets and
  high-dimensional matrices.
In {\em {HSCC}}, pages 41--50. {ACM}, 2018.
\href{http://dx.doi.org/10.1145/3178126.3178128}{doi:10.1145/3178126.3178128}.

\bibitem{dantzig1998linear}
George~Bernard Dantzig.
{\em Linear Programming and Extensions}, volume~48.
Princeton University Press, 1998.
\href{http://dx.doi.org/10.1515/9781400884179}{doi:10.1515/9781400884179}.

\bibitem{DunningHuchetteLubin2017}
Iain Dunning, Joey Huchette, and Miles Lubin.
{JuMP}: {A} {M}odeling {L}anguage for {M}athematical {O}ptimization.
{\em SIAM Review}, 59(2):295--320, 2017.
\href{http://dx.doi.org/10.1137/15M1020575}{doi:10.1137/15M1020575}.

\bibitem{featherstone2014rigid}
Roy Featherstone.
{\em Rigid body dynamics algorithms}.
Springer, 2014.

\bibitem{forets2021combining}
Marcelo Forets, Daniel~Freire Caporale, and Jorge M~P{\'e}rez Zerpa.
Combining set propagation with finite element methods for time integration in
  transient solid mechanics problems.
{\em Computers \& Structures}, 259:106699, 2022.
\href{http://dx.doi.org/10.1016/j.compstruc.2021.106699}{doi:10.1016/j.compstruc.2021.106699}.

\bibitem{forets2020efficient}
Marcelo Forets, Daniel Freire, and Christian Schilling.
Efficient reachability analysis of parametric linear hybrid systems with
  time-triggered transitions.
In {\em {MEMOCODE}}, pages 1--6. {IEEE}, 2020.
\href{http://dx.doi.org/10.1109/MEMOCODE51338.2020.9314994}{doi:10.1109/MEMOCODE51338.2020.9314994}.

\bibitem{ForetsS21}
Marcelo Forets and Christian Schilling.
Conservative time discretization: {A} comparative study.
{\em CoRR}, abs/2111.01454, 2021.

\bibitem{forets2021reachability}
Marcelo Forets and Christian Schilling.
Reachability of weakly nonlinear systems using {C}arleman linearization.
In {\em {RP}}, volume 13035 of {\em LNCS}, pages 85--99, 2021.
\href{http://dx.doi.org/10.1007/978-3-030-89716-1\_6}{doi:10.1007/978-3-030-89716-1\_6}.

\bibitem{frehse2011spaceex}
Goran Frehse, Colas~Le Guernic, Alexandre Donz{\'{e}}, Scott Cotton, Rajarshi
  Ray, Olivier Lebeltel, Rodolfo Ripado, Antoine Girard, Thao Dang, and Oded
  Maler.
{SpaceEx}: Scalable verification of hybrid systems.
In {\em {CAV}}, volume 6806 of {\em LNCS}, pages 379--395. Springer, 2011.
\href{http://dx.doi.org/10.1007/978-3-642-22110-1\_30}{doi:10.1007/978-3-642-22110-1\_30}.

\bibitem{ARCHCOMP20NONLINEAR}
Luca Geretti, Julien Alexandre~Dit Sandretto, Matthias Althoff, Luis Benet,
  Alexandre Chapoutot, Xin Chen, Pieter Collins, Marcelo Forets, Daniel Freire,
  Fabian Immler, Niklas Kochdumper, David~P. Sanders, and Christian Schilling.
{ARCH-COMP20} category report: Continuous and hybrid systems with nonlinear
  dynamics.
In {\em {ARCH}}, volume~74 of {\em EPiC Series in Computing}, pages 49--75.
  EasyChair, 2020.
\href{http://dx.doi.org/10.29007/zkf6}{doi:10.29007/zkf6}.

\bibitem{kamenev1996algorithm}
George~K Kamenev.
An algorithm for approximating polyhedra.
{\em Computational mathematics and mathematical physics}, 4(36):533--544, 1996.

\bibitem{kochenderfer2019algorithms}
Mykel~J Kochenderfer and Tim~A Wheeler.
{\em Algorithms for optimization}.
Mit Press, 2019.

\bibitem{kurzhanskiy2006ellipsoidal}
Alexander~A. Kurzhanskiy and Pravin Varaiya.
Ellipsoidal toolbox (et).
In {\em {CDC}}, pages 1498--1503, 2006.
\href{http://dx.doi.org/10.1109/CDC.2006.377036}{doi:10.1109/CDC.2006.377036}.

\bibitem{LeGuernic09}
Colas {Le Guernic}.
{\em Reachability analysis of hybrid systems with linear continuous dynamics}.
PhD thesis, Universit{\'e} Grenoble 1 - Joseph Fourier, 2009.

\bibitem{MathOptInterface-2021}
Beno{\^i}t Legat, Oscar Dowson, Joaquim Dias~Garcia, and Miles Lubin.
{M}ath{O}pt{I}nterface: a data structure for mathematical optimization
  problems.
{\em INFORMS Journal on Computing}, 2021.
\href{http://dx.doi.org/10.1287/ijoc.2021.1067}{doi:10.1287/ijoc.2021.1067}.

\bibitem{legat2021polyhedra}
Benoît Legat, Robin Deits, Gustavo Goretkin, Twan Koolen, Joey Huchette,
  Daisuke Oyama, and Marcelo Forets.
{J}ulia{P}olyhedra/{P}olyhedra.jl: v0.6.16, June 2021.
\href{http://dx.doi.org/10.5281/zenodo.4993670}{doi:10.5281/zenodo.4993670}.

\bibitem{LiuALSBK21}
Changliu Liu, Tomer Arnon, Christopher Lazarus, Christopher~A. Strong, Clark~W.
  Barrett, and Mykel~J. Kochenderfer.
Algorithms for verifying deep neural networks.
{\em Found. Trends Optim.}, 4(3-4):244--404, 2021.
\href{http://dx.doi.org/10.1561/2400000035}{doi:10.1561/2400000035}.

\bibitem{lotov2008modified}
Alexander~Vladimirovich Lotov and Alexis~I Pospelov.
The modified method of refined bounds for polyhedral approximation of convex
  polytopes.
{\em Computational Mathematics and Mathematical Physics}, 48(6):933--941, 2008.
\href{http://dx.doi.org/10.1134/S0965542508060055}{doi:10.1134/S0965542508060055}.

\bibitem{schilling2021verification}
Christian Schilling, Marcelo Forets, and Sebastian Guadalupe.
Verification of neural-network control systems by integrating {T}aylor models
  and zonotopes.
In {\em AAAI}, 2022.

\bibitem{SchuppAMK17}
Stefan Schupp, Erika {\'{A}}brah{\'{a}}m, Ibtissem~Ben Makhlouf, and Stefan
  Kowalewski.
{HyPro}: {A} {C}++ library of state set representations for hybrid systems
  reachability analysis.
In {\em {NFM}}, volume 10227 of {\em LNCS}, pages 288--294, 2017.
\href{http://dx.doi.org/10.1007/978-3-319-57288-8\_20}{doi:10.1007/978-3-319-57288-8\_20}.

\bibitem{sidrane2021overt}
Chelsea Sidrane, Amir Maleki, Ahmed Irfan, and Mykel~J. Kochenderfer.
{OVERT:} an algorithm for safety verification of neural network control
  policies for nonlinear systems.
\url{https://arxiv.org/abs/2108.01220}, 2021.

\bibitem{skogestad2007multivariable}
Sigurd Skogestad and Ian Postlethwaite.
{\em {M}ultivariable feedback control: analysis and design}, volume~2.
John Wiley \& Sons, 2007.

\bibitem{soto2021synthesis}
Miriam~Garc{\'{\i}}a Soto, Thomas~A. Henzinger, and Christian Schilling.
Synthesis of hybrid automata with affine dynamics from time-series data.
In {\em {HSCC}}, pages 2:1--2:11. {ACM}, 2021.
\href{http://dx.doi.org/10.1145/3447928.3456704}{doi:10.1145/3447928.3456704}.

\end{thebibliography}

\appendix
\section{How to install LazySets.jl}\label{sec:installation}

To use LazySets, first install Julia version v1.3 or higher\footnote{Julia binaries can be downloaded from \href{the official website}{https://julialang.org/downloads/}}. LazySets is a registered Julia package, and as such, you can install it by activating the \code{pkg} mode (type \code{]}, and to leave it, type \code{<backspace>}),
followed by

\begin{minipage}{\linewidth}
\vspace{-\abovedisplayskip}
\begin{lstlisting}
pkg> add LazySets
\end{lstlisting}
\end{minipage}
To load the package in a Julia session, do \code{using}, e.g.

\begin{minipage}{\linewidth}
	\vspace{-\abovedisplayskip}
	\begin{lstlisting}
julia> using LazySets

julia> HalfSpace([1.0, 0.0], 1.0)
HalfSpace{Float64, Vector{Float64}}([1.0, 0.0], 1.0)
	\end{lstlisting}
\end{minipage}

The LazySets reference manual is available online at
\href{https://juliareach.github.io/LazySets.jl/dev/}{https://juliareach.github.io/LazySets.jl/dev/}.

\section{Mathematical definitions}\label{sec:mathdef}

\subsection{Compact convex sets}\label{sec:convexdef}

Given a set $\X \subseteq \R^n$, its \emph{convex hull} is defined as
\[
	\CH(\X) = \left\{ \lambda \cdot x + (1 - \lambda) \cdot y ~\middle|~ x, y \in \X, \lambda \in [0, 1] \subseteq \R \right\}.
\]

A set $\X$ is \emph{convex} if it coincides with its convex hull.
A set is \emph{closed} if it contains all its boundary points.
A set $\X$ is \emph{bounded} if there exists a $\delta \in \R$ such that for all $x, y \in \X$ it holds that $\Vert x - y \Vert \leq \delta$.
A set is \emph{compact} if it is closed and bounded.

\subsection{Set operations}\label{sec:setops}

Given two sets $\X, \Y \subseteq \R^n$, the \emph{Minkowski sum} is
\[
	\X \oplus \Y = \{ x + y \mid x \in \X, y \in \Y \}
\]

The \emph{symmetric interval hull} of $\X$ is the smallest hyperrectangle that is centrally symmetric in the origin and contains $\X$.

Maps such as the \emph{linear map} $A \X$ are applied element-wise:
\[
	A \X = \{ A x \mid x \in \X \}
\]

\subsection{Basic properties of support functions}\label{sec:supfunc_properties}

The support function is a basic notion for approximating convex sets. Let $\X \subset \R^n$ be a compact convex set. The support function of $\X$ is the function defined as
\begin{equation}\label{eq:supfuncdef}
\rho(d, \X) := \max\limits_{x \in \X} d^\mathrm{T} x.
\end{equation}

We recall the following elementary properties of the support function.
Let $(d_1, d_2)$ denote the concatenation of vectors $d_1$ and $d_2$.
For all compact convex sets $\X, \Y$ in $\R^n$, $\Z$ in $\R^m$, all $n\times n$ real
matrices $M$, all scalars $\lambda$, and all vectors
$d, d_1 \in \R^n$, $d_2 \in \R^m$ we have:
\begin{align*}
    &\rho(d, \X \oplus \Y) = \rho(d, \X) + \rho(d, \Y) \\
    &\rho((d_1, d_2), \X \times \Z) = \rho(d_1, \X) + \rho(d_2, \Z) \\
    &\rho(d, \CH(\X \cup \Y)) = \max(\rho(d, \X), \rho(d, \Y)) \\
    &\rho(d, M \X) = \rho(M^T d, \X) \\
    &\rho(d, \lambda \X) = \rho(\lambda d, \X)
\end{align*}
Properties of the support vector (maximizers of \eqref{eq:supfuncdef}) can be found on the LazySets online documentation.
Analytic formulas for many important set types are known, allowing for efficient evaluations.
The LazySets docstrings contain mathematical explanations and references to the relevant literature (see for example \code{?Zonotope}).

\section{Code used in examples} \label{sec:code_examples}

The complete code for all examples can be found in the repository \url{http://github.com/JuliaReach/LazySets-JuliaCon21}. In this section we comment on some aspects of the code.

\subsection{Code for Fig.~\ref{sec:composition}} \label{sec:omega0}

The set $\X_0$ is a ball in the infinity norm of radius $0.1$ centered in $[1, 0]$, the set $E_+$ is a hyperrectangle centered in the origin, and $\Phi$ is a $2 \times 2$ matrix defined below.
In the context of reachability analysis for linear differential equations \cite{BogomolovFFVPS18}, the set $\X_0$ corresponds to the initial states, $E_+$ accouts for bloating terms, and $\Phi = e^{A\delta}$ is the state-transition matrix for some matrix $A$ and time step $\delta > 0$.

\begin{minipage}{\linewidth}
\vspace{-\abovedisplayskip}
\begin{lstlisting}
A = [0 1; -(4π)^2 0]
X₀ = BallInf([1.0, 0.0], 0.1)
δ = 0.025
Φ = exp(A*δ)
  2×2 Matrix{Float64}:
    0.95105652  0.02459079
   -3.88322208  0.95105652

r = [0.05477208, 0.07676220]
E₊ = Hyperrectangle(zeros(2), r)
Ω₀ = CH(X₀, Φ*X₀ ⊕ E₊)
\end{lstlisting}
\end{minipage}

\subsection{Code for Fig.~\ref{fig:supfunc}}

In Fig.~\ref{fig:supfunc}, the set $\X$ is a polygon in vertex representation. Such a \code{VPolygon} can be constructed from a vector of points, or simply a matrix where each column corresponds to the coordinates of a point. (For higher-dimensional sets in vertex representation the set type \code{VPolytope} is used).

\begin{minipage}{\linewidth}
\vspace{-\abovedisplayskip}
\begin{lstlisting}
# two-dimensional polygon in vertex representation
X = VPolygon([-3   -2   0   1  2  0  -0.8;
              0.6  -2  -2  -1  1  2   1.8])
\end{lstlisting}
\end{minipage}

The supporting half-space of $\X$ is computed by evaluation of the support function along the direction of interest.

\begin{minipage}{\linewidth}
\vspace{-\abovedisplayskip}
\begin{lstlisting}
# computing a supporting half-space
d = [-1.0, 1.0]
sf = ρ(d, X)
H = HalfSpace(d, sf)
\end{lstlisting}
\end{minipage}

\subsection{Code for Fig.~\ref{fig:overapproximate}}

The set $\X$ is the same as in Fig.~\ref{fig:supfunc}.
We overapproximate it with a box (\code{Y}) and two zonotopes (\code{Z} and \code{W}).

\begin{minipage}{\linewidth}
\vspace{-\abovedisplayskip}
\begin{lstlisting}
Y = box_approximation(X)
Z = overapproximate(X, Zonotope, PolarDirections(3))
W = overapproximate(X, Zonotope, PolarDirections(5))
\end{lstlisting}
\end{minipage}

The third argument to \code{overapproximate} here represents a list of vectors that are used to synthesize the generators of the resulting zonotope.
The type \code{PolarDirections} lazily represents vectors that uniformly cover the unit disc, starting with the vector $(1, 0)^T$.

\begin{minipage}{\linewidth}
\vspace{-\abovedisplayskip}
\begin{lstlisting}
collect(PolarDirections(5))
5-element Vector{Vector{Float64}}:
 [1.0, 0.0]
 [0.30901699437494745, 0.9510565162951535]
 [-0.8090169943749473, 0.5877852522924732]
 [-0.8090169943749475, -0.587785252292473]
 [0.30901699437494723, -0.9510565162951536]
\end{lstlisting}
\end{minipage}

\subsection{Code for Section~\ref{sec:taylormodels}} \label{sec:taylormodels_appendix}

The linear Taylor models are $p_1(t) = 2 + t$ and $p_2(t) = 0.9 + 3t$ with remainders $I_1 = I_2 = [-0.5, 0.5]$ in the domain $D = [-3, 1]$ and centered at zero.
The non-linear case has $p_1(t)$ as in the linear one and $p_2(t) = 0.9 + 3t + 3t^2$ with the same remainders and domain as in the linear case.

\begin{minipage}{\linewidth}
\vspace{-\abovedisplayskip}
\begin{lstlisting}
using TaylorModels
const IA = IntervalArithmetic
const TM1 = TaylorModel1

I = IA.Interval(-0.5, 0.5)
x₀ = IA.Interval(0.0)
D = IA.Interval(-3.0, 1.0)
p1 = Taylor1([2.0, 1.0], 2)
p2 = Taylor1([0.9, 3.0], 2)

# define vector of linear Taylor models
vTMlin = [TM1(pi, I, x₀, D) for pi in [p1,p2]]

# define vector of non-linear Taylor models
p2nl = Taylor1([0.9, 3.0, 3.0], 3)
vTMnonlin = [TM1(pi, I, x₀, D) for pi in [p1, p2nl]]
\end{lstlisting}
\end{minipage}

\subsection{Code for Section~\ref{sec:custom_arrays}}

Special array operations and the type \code{SingleEntryVector} are available in the LazySets submodule \code{LazySets.Arrays}. This example consists of the concrete intersection of two 10-dimensional sets, one of them which is unbounded (a half-space).

\begin{minipage}{\linewidth}
\vspace{-\abovedisplayskip}
\begin{lstlisting}
const SEV = LazySets.Arrays.SingleEntryVector

X = Hyperrectangle(zeros(10), 2*ones(10))
Hsev = HalfSpace(SEV(1, 10, 1.0)), 1.0)
Hvec = HalfSpace(Vector(Hsev.a), 1.0)
\end{lstlisting}
\end{minipage}

\end{document}